\documentclass[twocolumn]{openjournal}

\usepackage[T1]{fontenc}
\usepackage{ae,aecompl}
\usepackage{natbib}
\usepackage{wasysym}

\usepackage{xcolor}
\begin{document}
\title{Bound circumplanetary orbits under the influence of radiation pressure: Application to dust in directly imaged exoplanet systems} 
\author{Brad M. S. Hansen}
\email[email]{ hansen@astro.ucla.edu}
\affiliation{ Mani L. Bhaumik Institute for Theoretical Physics,  Department of Physics \& Astronomy, University of California Los Angeles, \\ Los Angeles, CA 90095}
\author{ Kevin Hayakawa}
\affiliation{
California State University Channel Islands, 1 University Dr, Camarillo, CA 93012}



\begin{abstract}
We examine the population of simple periodic orbits in the Hill problem with radiation pressure included, in order to understand the distribution of gravitationally bound dust in orbit around a planet. We study a wide range of radiation
pressure strengths, which requires the inclusion of additional terms beyond those discussed in previous analyses 
of this problem. In particular, our solutions reveal two distinct populations of  stable wide, retrograde, orbits, as opposed to
the single family that exists in the purely gravitational problem.

We use the result of these calculations
to study the observational shape of dust populations bound to extrasolar planets,  that might be observable in scattered
or reradiated light. In
particular, we find that such dusty clouds should be elongated along the star--planet axis, and that the elongation of the bound population increases with $\beta$, a measure of the strength of the radiation pressure.

As an application of this model, we consider the properties of the Fomalhaut system. The unusual orbital properties of
the object Fomalhaut~b can be explained if the observed light was scattered by dust that was released from an object in a quasi-satellite orbit about a planet located in, or near, the observed debris ring. Within the context of the model of Hayakawa \& Hansen (2023), we find that the dust cloud around such a planet  is still approximately an order of magnitude fainter than
the limits set by current JWST data.

\end{abstract}

\maketitle


\section{Introduction}

The late stages of planetary system assembly  are expected to result in the production of copious amounts
of dust, which can be observed due to its capacity for reprocessing the light from the central star. Imaging of
this population of dust, either in scattered light or thermal emission, can provide information on the properties
of the planetary system by virtue of the sensitivity of the dust to the gravitational influence of the planets
in the system.

In \cite{HH23}, we present a model for the origin of thin dust rings, in which the dust is generated in  irregular satellite systems.
As an alternative to the common `birth ring' model for dust populations -- which posits an origin in a ring of
colliding planetesimals -- we invoke a population of irregular satellites that generates the dust in the local
vicinity of the host planet. The effect of radiation pressure from the central star is to shift the pseudopotential so
that the collisional cascade is halted when dust escapes via the $L_2$ point and goes into orbit exterior to the planet. We demonstrated
that this dust, regulated by the contours of the pseudopotential, naturally yields a thin dust ring similar to those observed
in some systems.

Not all of the dust immediately escapes to orbits exterior to the planet. The question we wish to address in this
paper is whether other aspects of the model are amenable to observation. In particular, we wish to calculate 
the expected properties of dust trapped in stable orbits, and whether they can be observed.  Perhaps the
best known example of a thin ring system is that in orbit around the star Fomalhaut \citep{Kalas05}. There have also been
claims of a planetary object in this system \citep{Kalas08,Kalas13} although this has also been argued to be a dust cloud not
associated with a planet \citep{Currie12,Gal13}. The optical colors of this object suggest that the origin of the emission is
scattered light from the primary, and \cite{KW11} have suggested that the emission is the result of the collisional grinding down of a population of irregular satellites as posited above. 

With the continued imaging of dust systems with HST, and the new capabilities of JWST becoming available, it is timely
to reconsider the observability of dust bound, or in close orbit, around a planet, and to evaluate the expected signatures
of this dust. \cite{KW11} assume a distribution of dust that follows the orbits of the parent bodies about the planet, but
particles in orbit around a planet will be subject to a variety of forces due to the radiation from the central star \citep{Burns79}.
Thus, in \S~\ref{PhotoHen} we define the version of the Photogravitational Hill problem needed for our case.
In \S~\ref{PerOrb}  we  then generalise the families of periodic  orbits discussed in the original Hill problem by \cite{Hen69,Hen70}. In \S~\ref{Obs}
we then discuss how these orbits will manifest themselves in the full restricted three-body co-ordinate system and how this might be observable.

\section{The Photogravitational Hill problem and Henon orbit families}
\label{PhotoHen}

The motion of dust in stellar or planetary systems is well suited to description in the limit of the restricted three-body problem, as the mass of dust particles is so small as to be accurately described as a test particle. In the case where the two massive bodies have a circular orbit, the dynamics is well-known and regulated by a conserved integral, the Jacobi integral \cite[e.g.][]{MD2000}.

In the limit where the mass ratio $\mu$ between the two  massive bodies is small, the dynamics in and around the sphere of influence of the
less massive body can be fruitfully described in the context of the Hill problem \citep{Hill78a,Hill78b,Hill78c}, wherein the dynamics is described in
a co-ordinate frame centered on the smaller body (the planet). \cite{Hen69} presented an elegant analysis of the
simply periodic orbits in a rescaled version  of Hill's problem in the limit where $\mu \rightarrow 0$. We wish here to revisit this
question when the test particles are subject to a radiation pressure force from the central star, in order to describe the kinds of structures we might observe in a population of dust particles generated in close proximity to a planet.

In the original restricted three body problem, the massive bodies are separated by a semi-major axis of unity, and the center
of mass in the co-rotating frame lies at $(0,0,0)$.  If we wish to describe the motion about the planet, it is helpful to 
move the origin of the co-ordinate system to the position of the
planet (at $x=1-\mu$ and $y=0$).  \cite{Hen69,Hen70} noticed that one could describe the dynamics quite generally by scaling out the mass ratio in the definition of the circumplanetary co-ordinates $\xi$, $\eta$ and $\zeta$. We follow this approach, resulting
in the circumstellar-to-circumplanetary coordinate transformation 
$x =  1 - \mu + \mu^{1/3} \xi$, $y=\mu^{1/3} \eta$ and $z=\mu^{1/3} \zeta$.
 The effect of the radiation pressure is to reduce the effective gravity of the central, luminous, object by a factor $(1 - \beta)$, where $\beta$ parameterises the strength of the radiation pressure.
 The resulting dynamics of the test particle is given by the equations
\begin{eqnarray}
\ddot{\xi} - 2 \dot{\eta} & = &  \frac{\beta}{\mu^{1/3}} + \left( 3 - 2 \beta  - \frac{1}{\Delta^3}\right) \xi \label{Photo1} \\
\ddot{\eta} +  2 \dot{\xi} & = & \left( \beta - \frac{1}{\Delta^3} \right) \eta \label{Photo2} \\
\ddot{\zeta} & = & \left( \beta -1 - \frac{1}{\Delta^3} \right) \zeta \label{Photo3}
\end{eqnarray}
where $\Delta^2 = \xi^2 + \eta^2+\zeta^2$ and we have dropped terms of higher order than $\mu^{1/3}$.
 We have also restricted ourselves to the radial component of the radiation pressure force. The scattering of light by an object in motion around the source also results in an azimuthal component, which gives rise to the Poynting-Robertson drag. However, this force is smaller than the radial component \citep{Burns79} by a factor $\sim v/c$ (where $v$ is the particle's orbital motion). For objects on scales $\sim 10$--100~AU, this is $< 3 \times 10^{-5}$. This is much smaller than terms of order $\mu^{1/3}$ and thus is neglected here.

The dynamics of small particles in the restricted three body problem with radiation pressure has been described in many
prior studies 
\citep[][e.g]{Rad50,Rad53,Sch80,SMB85,Kush08,Zot15}.
These analyses describe how the nature and stability of the Lagrangian equilibria evolve as the balance between gravity and radiation changes. Of particular interest to us are studies such as those by \cite{MRV00,KMP02,PPK12},  which seek to describe the dynamics of particles near a planet while subject to radiation pressure, and which catalog the kinds of simply periodic orbits that arise.

However, these prior analyses of the `photogravitational problem' make an approximation that is not appropriate for the
application we seek. 
After the derivation of the above equations,  the authors  next step is to set $\beta = \mu^{1/3} Q_1$
and then take the limit $\mu \rightarrow 0$. This achieves an elegant rescaling  that removes the mass dependance, in the same
manner as the original classic work of Henon. However, it comes at a cost -- in order for $Q_1$ to be a constant, $\beta \rightarrow 0$ in lockstep with $\mu$, so
this only applies in the limit of small radiation pressure. This may be appropriate to applications such as describing the slow drift of spacecraft under the influence of stellar radiation \citep[e.g.][]{GCT14,YSM15}, but it is not appropriate for describing the dynamics of dust, where
the value of $\beta$ need not be infinitesimal.

Thus, we opt to keep $\beta$ finite, so that we may retain our ability to describe the dynamics of dust experiencing significant
perturbations due to radiation pressure. This comes at the price that our description is no longer scale free -- for a given $\beta$
we must also specify the value of $\mu$.
We will, however, 
 denote $Q'=\beta/\mu^{1/3}$, as prior authors have done, to enable direct comparison between their results and ours.

The equations (\ref{Photo1}) and (\ref{Photo2}) still admit an equivalent of the Jacobi integral, now given as
\begin{equation}
C_H = \left( 3 - 2 \beta \right)\xi^2 + \beta \eta^2 + \left( \beta - 1 \right) \zeta^2 + \frac{2}{\Delta} + 2 Q' \xi - \dot{\xi}^2 - \dot{\eta}^2 - \dot{\zeta}^2
\end{equation}

The difference between these equations and those of the traditional `photogravitational Hill problem' \citep{MRV00,KMP02}   leads to some qualitative differences, as we will see below (although they represent a subset of the
equilibria discussed in the more general restricted three body problem -- \cite{SMB85}).

\subsection{The Planar Equilibrium Points}

The equilibrium points of the equations for the restricted three body problem define the transitions between different classes of dynamics. 
In the classic reduced three body problem, there are five equilibrium points -- the Lagrange equilibria.
Three of them -- the colinear points $L_1$, $L_2$ and $L_3$ -- lie on the line between the two gravitating masses in the co-rotating
frame. The other two -- the triangular points $L_4$ and $L_5$ -- lie on either side of the secondary, at angles $\pm 60^{\circ}$.

The introduction 
of the radiation force shifts the positions of the equilibria and even introduces a new set of out-of-plane equilibria in the case of very strong radiation pressure \citep{Rad50,Rad53,Sch80,SMB85}. In order to understand the dynamics of dust particles, we wish to understand the implications of these changes in the Hill limit (i.e. when $\mu \ll 1$).

\subsubsection{The Colinear Points}

In the purely gravitational case ($\beta = 0$), the equations in the Hill limit require $\eta = 0$ for equilibria to exist,
 which implies $\Delta = \left| \xi \right|$.
When we feed this condition into 
equation~(\ref{Photo1}), we find that the colinear equilibria must satisfy
\begin{equation}
\left|\xi\right|^2 \xi + \frac{Q'}{3 - 2 \beta} \left|\xi\right|^2 - \frac{1}{3 - 2 \beta} \frac{\xi}{\left| \xi \right|} = 0, \label{Cubic}
\end{equation}
which can be written as two different cubic equations for positive and negative $\xi$, with different signs for the second term.

In the case of $\beta=Q'=0$, this yields the usual criterion $\xi = \pm 3^{-1/3}$. Thus, the $L_1$ (negative $\xi$)
and $L_2$ (positive $\xi$) equilibria lie at the same distance from the planet, on either side. The distant $L_3$
equilibrium vanishes in the limit.
 For finite $\beta$, there are two
different cubic equations, depending on the sign of the $Q'$ term in equation~(\ref{Cubic}), which will
yield different values of $\left| \xi \right|$ on either side of the planet. Although cubic equations can yield multiple solutions, we note that not all of the solutions
to equation~(\ref{Cubic}) are valid, because  valid solutions must have the correct relationship between  $\xi$
and $\left| \xi \right|$ as determined by the sign of the $Q'$ term.
 The result of this is that the $L_1$ and $L_2$ points now lie at different distances from the planet.
As we increase $\beta$, the $L_1$ point moves closer to the star and the $L_2$ point moves closer to the planet.
 The $L_3$ point also remains absent in this case. The character of the
equipotentials also changes with finite $\beta$. Interior to $L_2$, the equipotentials remain centered on the planet, and so
the zero-velocity contours restrict particles to circumplanetary motion. However, between the contour passing through $L_2$ and that passing through $L_1$, the zero-velocity curves open up and allow particles to escape to circumstellar orbits
(this is the essential element that allows narrow dust rings to form in the model of \cite{HH23}). As a result, the range of
circumplanetary orbits is not bounded by $L_1$ and $L_2$ but by $L_2$ and a location $L_2'$, representing the
innermost edge of the pseudopotential contour that passes through  $L_2$.

\begin{figure}
\centering
\includegraphics[width=1.0\linewidth]{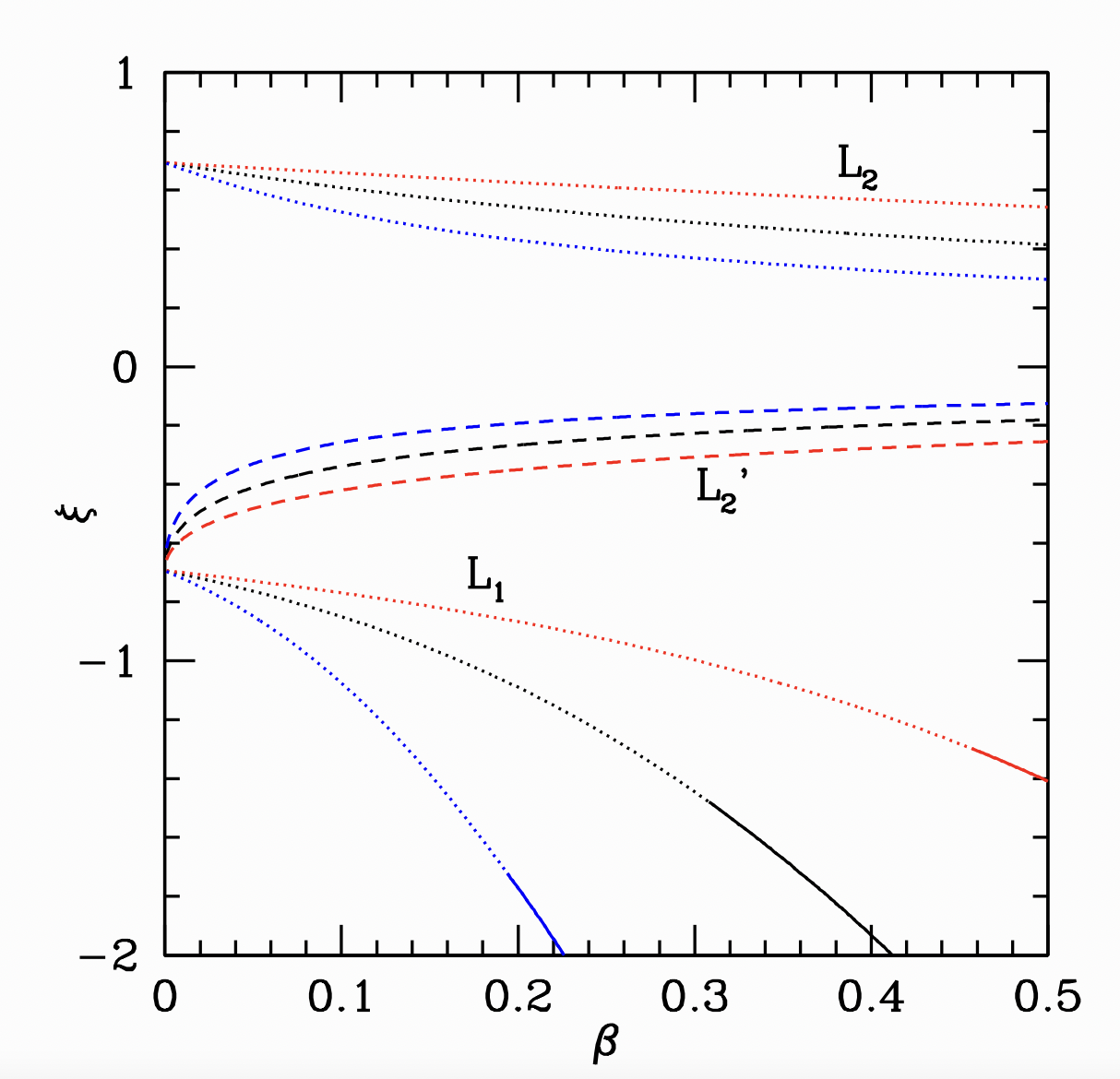}
\caption{The dotted lines indicate the position of the saddle points $L_1$ and $L_2$, for the
stated value of $\beta$. The $L_1$ curve turns solid at the point that it transitions from a saddle point to a potential minimum. The dashed contour indicates the location $L_2'$ that marks the inner edge of the pseudopotential that passes through the $L_2$ point. The blue contours are for a mass ratio of $\mu=10^{-4}$, the black contours are for $\mu=10^{-3}$ and the red contours are for $\mu=10^{-2}$. }
 \label{Colins}
\end{figure}

 The nature of the equilibria will depend on the second derivatives of the pseudopotential. At low values of $\beta$, both the
 $L_1$ and $L_2$ points remain saddle points, as in the purely gravitational case. However, as $\beta$ increases, the
 $L_1$ point becomes a potential minimum when $\xi_1 = - \beta^{-1/3}$ (this is when $\partial^2 C_H/\partial \eta^2$ changes sign.) Figure~\ref{Colins} shows how the locations of $L_1$, $L_2$ and $L_2'$ vary with the value of $\beta$ (for the case $\mu=0.001$).


\subsubsection{Triangular Points in the Hill Problem}

The presence of the $\beta$ dependant term in equation~(\ref{Photo2}) enables the existence of additional
equilibria beyond the colinear points.
The new solution condition is $\Delta = \beta^{-1/3}$, which amounts to a condition on a specified distance from the planet.
 If we put this into equation~(\ref{Photo1}) we must satisfy
\begin{equation}
\xi = - \frac{Q'}{3(1-\beta)} \label{Dbeta}
\end{equation}
This yields something very like the $L_4$ and $L_5$ points, which don't appear in the usual Hill or photogravitational Hill problems.
This is a consequence of the $\beta$ dependant terms that remain even after the Taylor expansion of the potential about the
location of the secondary. These terms break
 the usual symmetry that exists in the other versions of the Hill problem.
  As such, these points are not  really a qualitatively new equilibrium, but they will have an influence on the kinds of periodic orbits we seek to describe below.

\begin{figure}
\centering
\includegraphics[width=1.0\linewidth]{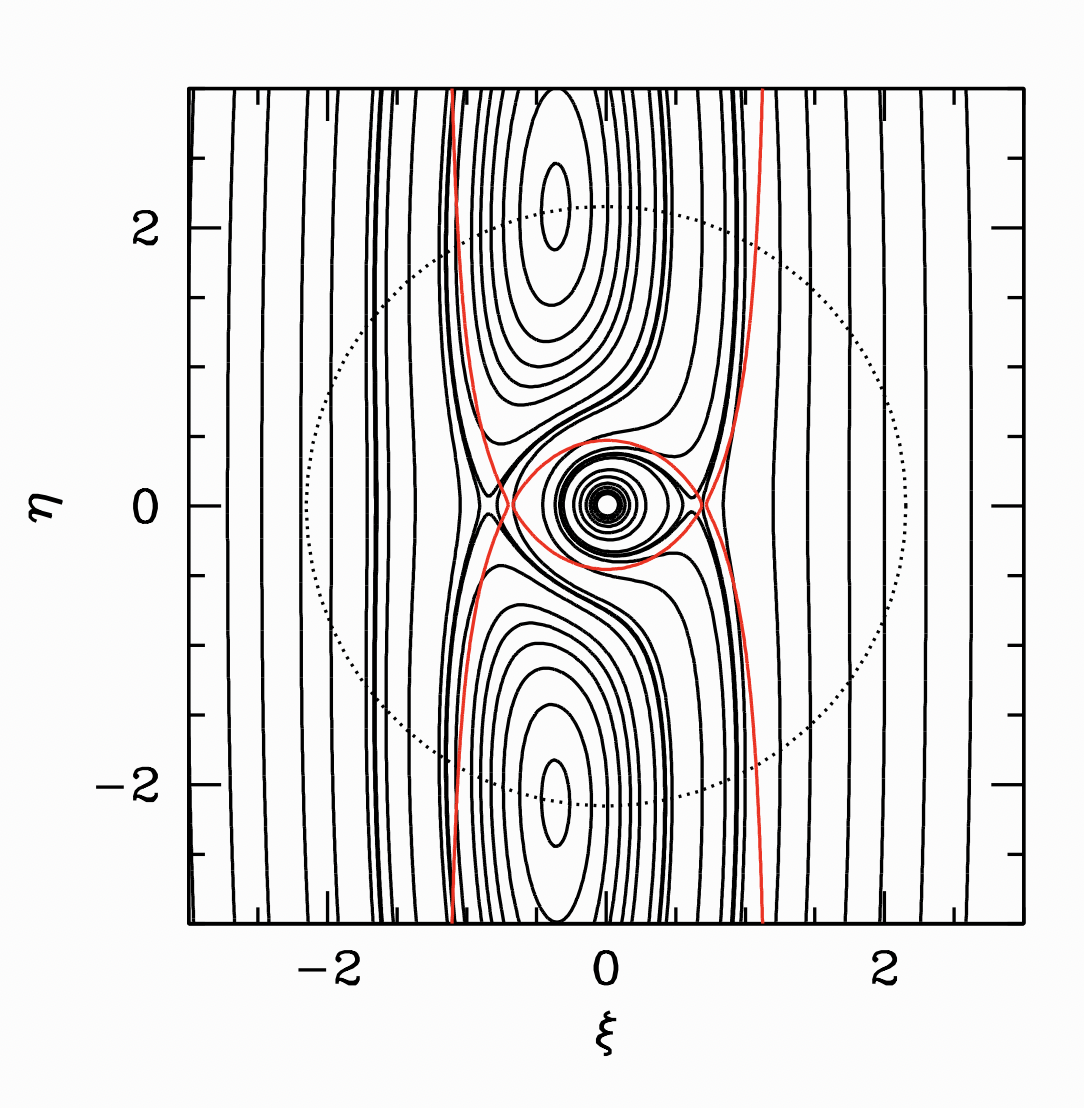}
\caption{The black contours show equipotentials for the case of $\mu=10^{-3}$ and $\beta=0.1$ (so $Q'=1$). The red contour shows the separatrix
for the case of $\beta=0$ (and the same mass). The dotted line represents the circle upon which the Triangular equilibria are expected, and we
can see the appearance of these points in the contours. }
 \label{Beta1}
\end{figure}

Figure~\ref{Beta1} shows an example for $\beta=0.1$ and $Q=1$ (so $\mu=10^{-3}$) compared to the $\beta=0$ case for the same mass (in red).  The red contour shows the seperatrix for the $\beta=0$ case, delineating the boundary between circumplanetary orbits, interior and exterior heliocentric orbits, and the forbidden region. The inclusion of finite radiation pressure introduces a new set of
solutions, 
 defined by the condition $\Delta = \beta^{-1/3}$ and shown by the dotted circle. This 
 leads to the solution 
defined by equation~(\ref{Dbeta}), resulting in the new $L_4$ and $L_5$ equilibria as shown by the new extrema in the solid black contours. We see that the circumplanetary region is now compressed and there is a region of mostly-exterior  heliocentric orbits that now pass interior to the planet briefly, between the $L_1$ point and the circumplanetary region.

We dub these points the Triangular points because of their obvious connection to the $L_4$ and $L_5$ equilibria in the traditional restricted three body problem, but we must also note the features that do not exhibit an exact correspondence. In particular the angle these equilibria make with the primary--secondary axis is not $60^{\circ}$, and depends on $\beta$. 
As $\beta$ increases, these points move closer to the axis, and will converge when $\beta^{-1/3} = Q'/3 (1-\beta)$. Since
$Q'=\beta/\mu^{1/3}$, we can expand in the small parameter $\mu$ to derive an approximate criterion
\begin{equation}
\beta_{con} \sim 2.279 \mu^{1/4} \left( 1- 1.709 \mu^{1/4} + 1.949 \mu^{1/2}\right)
\end{equation}
For $\mu = 10^{-3}$, this criterion yields $\beta_{con}=0.307$ (a  direct numerical solution yields the root $\beta_{con}=0.306$).

We also note that the criterion for $L_4$ and $L_5$ to merge on the axis also corresponds to the condition for the change 
of $L_1$ from a saddle point to a potential minimum. Note that, with our sign convention, this does not correspond
to a fixed point for stable orbits. As we shall see, $L_1$ remains a locus for unstable equilibria.

\subsection{Equilibria out of the plane}
\label{OutPlane1}

In the case of very strong radiation pressure, a  new set of equilibria is possible that simply do not
appear in the purely gravitational problem \citep{Sch80,SMB85}. This occurs if the radiation pressure
from one of the objects is sufficient to completely overwhelm the gravitational influence of that body.
In this case it is possible to achieve equilibria that lie out of the orbital plane.

In the context of our problem, this will only occur if $\beta >1$, in which case the right hand side of
equation~\ref{Photo3} can be set to zero if $\Delta = (\beta -1)^{-1/3}$, even if $\zeta \neq 0$. In the usual context of dust
in orbit around a star, $\beta>0.5$ leads to unbound trajectories, but the presence of the planetary
gravity may open up the possibility of equilibria featuring very small dust that feels strong radiation
pressure.

Further conditions on these equilibria are that $\eta = 0$ (from equation~\ref{Photo2})  -- so that
these equilibria lie in the plane passing through the planet and the star -- and that
\begin{equation}
\xi = - \frac{Q'}{4 - 3 \beta}.
\end{equation}
We also have that $\Delta^2 = \xi^2 + \zeta^2$ in this case, so that the requirement that $\zeta^2>0$ places
some restrictions on the solution, namely that 
\begin{equation}
Q' < \frac{ \left| 4 - 3 \beta \right|}{(\beta - 1)^{1/3}}.
\end{equation}
Since $Q' = \beta/\mu^{1/3}$, this places a restriction on the mass, namely
\begin{equation}
\mu > \beta^3 \frac{(\beta - 1)}{\left|4 - 3 \beta\right| ^3}
\end{equation}

\begin{figure}
\centering
\includegraphics[width=1.0\linewidth]{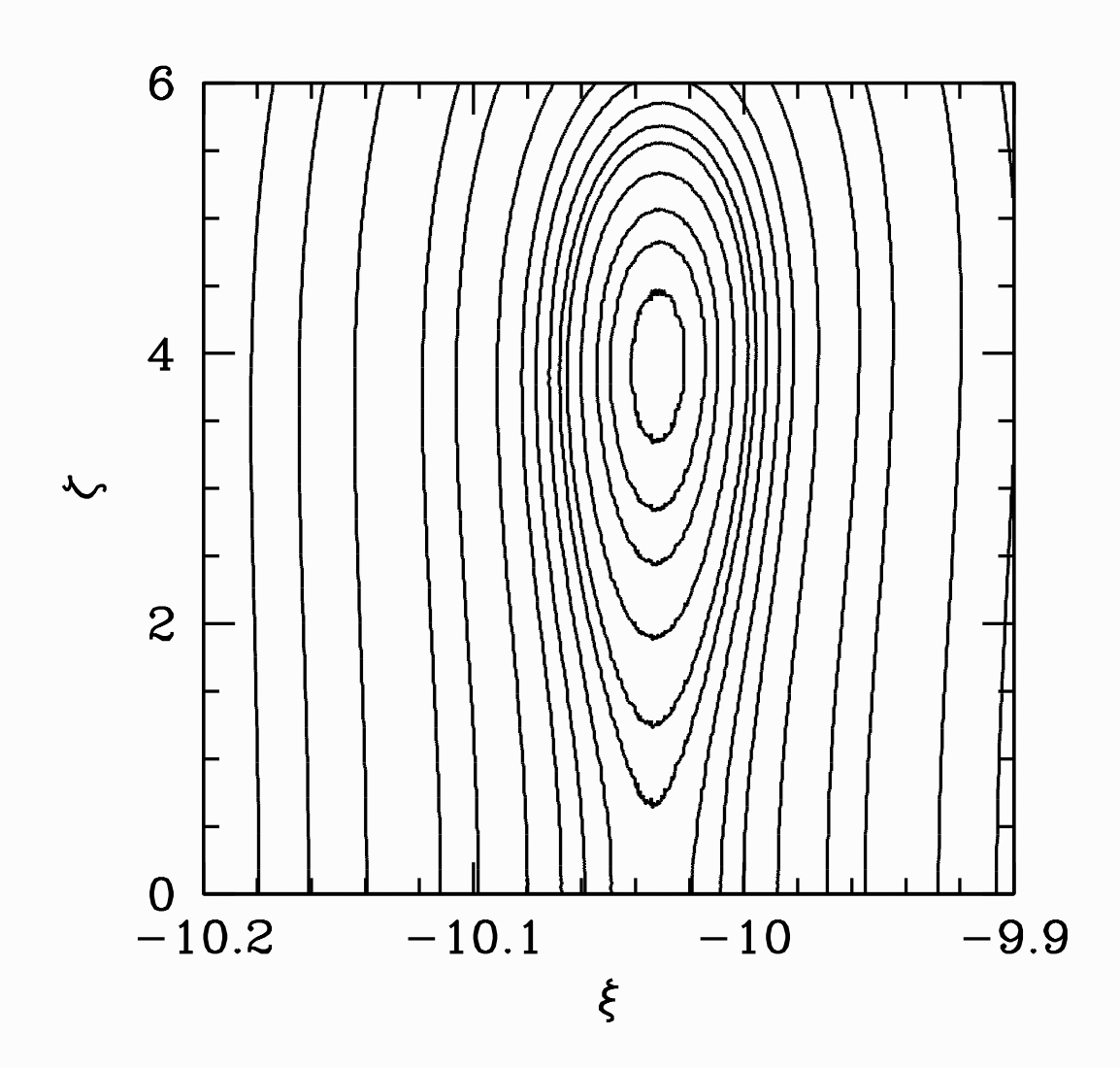}
\caption{The contours show equipotentials in the $\eta=0$ plane for the case $\mu=0.001$ and $\beta=1.0008$. We see that there is a narrow libration region about this equilibrium point, potentially available to very small dust whose radiation pressure can outweight the gravity of the host star.
 \label{Zeq}}
\end{figure}

Thus, for planetary mass ratios, such equilibria are possible, but only for particles whose $\beta$ is 
infinitesimaly above unity.  As an example,
Figure~\ref{Zeq} shows the  contours of the  potential about the equilibrium point at $\xi=-10.032$ and
$\zeta=3.924$ for the case of $\mu=0.001$ and $\beta=1.0008$.  Note that the range on the horizontal axis 
is quite small -- 
 the libration region is very narrow in
$\xi$, although it can extend over several scale heights in $\zeta$,  as demonstrated by the large range on
the vertical axis. Thus, these equilibria are narrowly confined to a essentially a line-segment above the plane, on the line between the planet and the star

Although these equilibria exist in the Hill expansion version, they occur only close to the primary (as the radiating body) and, as such, are 
not relevant to the question of dust structures in the vicinity of the secondary.

\section{Periodic Orbits}
\label{PerOrb}

Our principal goal is to understand the nature of long-lived dust particles orbiting a planet, such that they might be imaged in scattered light.  In order to better understand this, we wish to understand the orbits that are stable around the planet, subject to the combined effects of gravity and radiation pressure. \cite{Hen69,Hen70}  presented a classic analysis of the different families of periodic orbits in the Hill problem. We wish to understand how the effects of radiation pressure change this.
Before we do that, it is worth briefly reviewing the simply periodic orbit families as classified by Henon. 

Families $a$ and $c$ represent the (unstable) extensions of the libration orbits about the Lagrange equilibria at $L_1$ and $L_2$. 
The families $g$ and $f$ represent the prograde and retrograde satellite orbits of the planetary body. The prograde orbit family
splits, at critical point $g1$, with the introduction of a second orbital family $g'$ (which contains both a stable and unstable equilibrium for some ranges of $C_H$). Family $g$ also becomes doubly periodic at a critical point $g2$.

\subsection{Prograde Orbits}

A detailed survey of the simple periodic orbits requires direct numerical integration of the equations of
motion~(\ref{Photo1}),(\ref{Photo2}) and (\ref{Photo3}). Here we will restrict ourselves to the orbital plane, as the out-of-plane
equilibria are too far from the planet to be realistically included in the Hill limit.
We adopt a similar orbital classification procedure as discussed in \cite{Hen69}. We begin integrations with $\eta=0$ and $\dot{\xi}=0$
and choose an initial value for $\xi$, and a value for $\dot{\eta}$ based on an assumed value of $C_H$. The positive sign
of $\dot{\eta}$ is chosen and we record the $\xi$ and $\dot{\xi}$ every time the orbit crosses the $\eta=0$ plane moving in the positive direction. This is used to construct a Poincare plot. Equilibria are defined as those orbits for which $\xi$ does not change from the initial value on the first complete orbit (i.e. simple periodic orbits in the definition of Henon). The stability of
each equilibrium is evaluated by constructing the linear mapping of orbits surrounding the equilibrium and evaluating the eigenvalues of the resulting matrix \citep{Tremaine23}. We map out the families of simply periodic orbits and will also describe a subset of the doubly periodic orbits which
can impact the stability of some of the simply periodic families.

\begin{figure}
\centering
\includegraphics[width=1.0\linewidth]{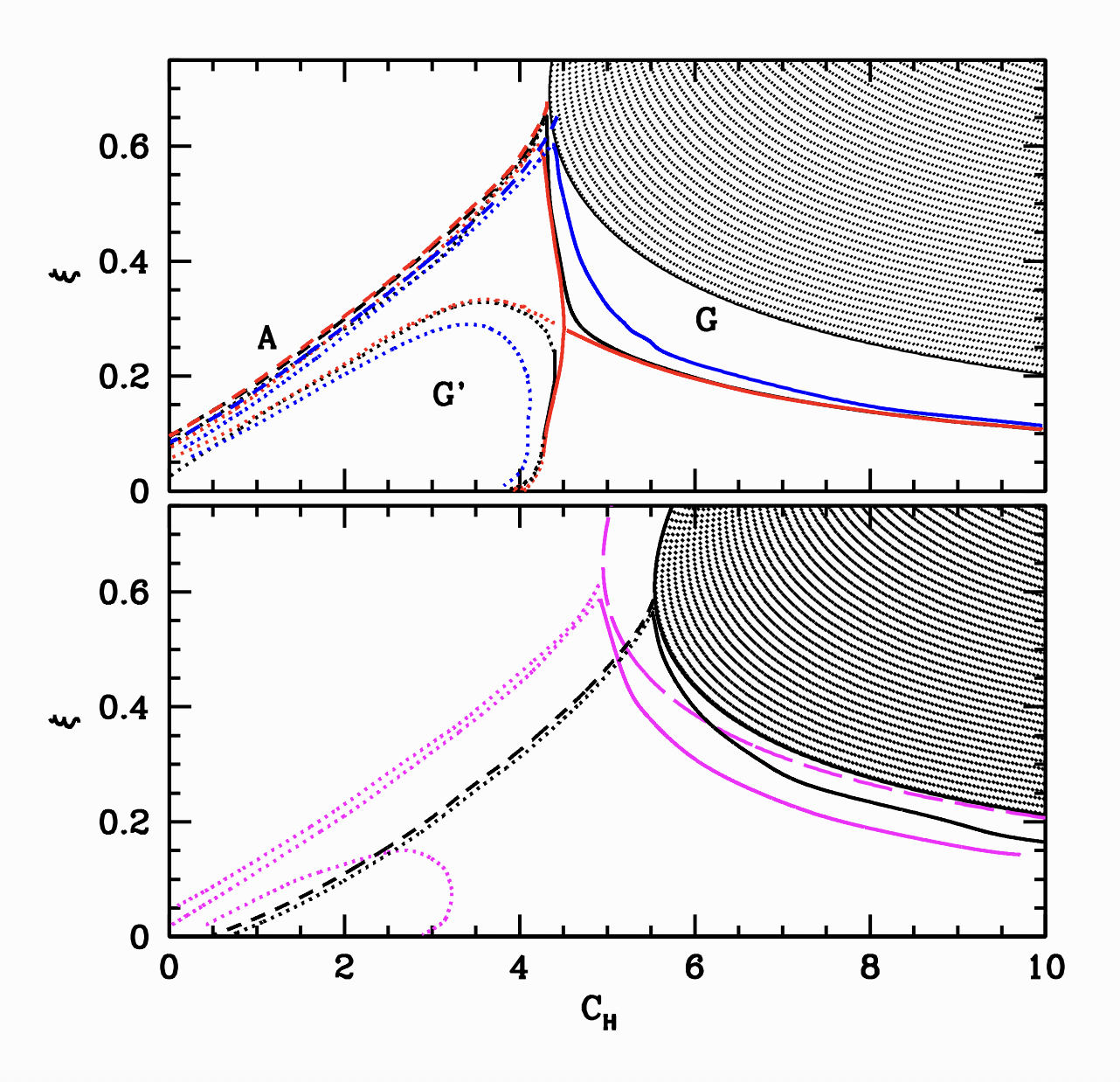}
\caption{The dark shaded region indicates the region in which orbits are  disallowed. In the upper panel this is shown
for $\beta=0$.
 The red curves  show the prograde equilibrium orbit families for the case $\beta=0$. The black curves show how these shift once a small radiation pressure
($\beta=0.001$, $Q'=0.02$) is included. The blue curve shows the case for $\beta=0.01$ ($Q'=0.2$).  In the lower 
panel, 
the magenta curves show the case for $\beta=0.05$ and the black curves show the case for $\beta=0.1$. The shaded  forbidden region
shown here is for $\beta=0.1$ and the magenta long dashed line shows the boundary of the forbidden region in the case
$\beta=0.05$. In both panels,
the solid lines
indicate those equilibria that are stable, and the dotted lines indicate the unstable portions of the $G/G'$ families. The
dashed lines indicate the unstable $A$ family.
 \label{Orbit1}}
\end{figure}

Figure~\ref{Orbit1} shows the resulting prograde equilibria, for the mass ratio $\mu=10^{-3}$.
In the upper panel, the red curves represent the original orbital families for $\beta=0$ from \cite{Hen69}.
The black curves  show the case for weak radiation pressure ($\beta=0.001$, so $Q=0.02$). The red curves show the  original
Henon orbit families of $g$ and $g'$ (and the unstable family $a$).  In the case of $\beta=0$, the appearance of family $g'$ occurs
at $C_H=4.4999$ and $\xi_0=0.2835$ and all three equilibria move away from this point as $C_H$ decreases.
 In the case of non-zero $\beta$ we find that, instead of a point of intersection, the new family $G'$ emerges at lower values
 of $\xi$ than the equilibrium on the $G$ family at the corresponding $C_H$. The evolution, in this case, is more akin to an avoided crossing of two families, 
  $G$ and $G'$, each of which contains a portion of
the original $g$ and $g'$. 
A similar behaviour  occurs in the prior definition of the photogravitational Hill problem \citep{MRV00,KMP02}, albeit with different $G$ and $G'$ (corresponding to linking different combinations of the sections of $g$ and $g'$). In our case, the family $G$ extends from high $C_H$ all the way to the Lagrange point, with the equilibria distorting from quasi-circular shapes (deep in the potential well) to more elongated shapes as they approach the Lagrange point.  The family $G'$, on the other hand, only exists at moderate $C_H$, linking a stable branch of compact eccentric orbits to the unstable branch that was originally the extension of the g family.

The blue curves in the upper panel of Figure~\ref{Orbit1} show the equilibria for the case of $\beta=0.01$. The $G$ family shifts to larger $C_H$ at fixed $\xi$, while
the $G'$ family shifts to the left (lower $C_H$ at fixed $\xi$). Note also that the $A$ family (the analog of the $a$ family) shifts down (to lower $\xi$ at
fixed $C_H$), so that the $A$ and $G'$ families start to approach each other as $\beta$ increases. 
The lower panel of Figure~\ref{Orbit1} shows the evolution of the orbital families as $\beta$ increases.
The magenta curves show the case of $\beta=0.05$. As the radiation pressure grows, the forbidden region moves to large $C_H$ and $G$ family
shifts in the same direction. Conversely, the $G'$ family shifts down to lower $C_H$. The black curves show the case
for $\beta=0.1$. In this case, the family $G'$ no longer exists.

In the case of $\beta=0$, the association of the two branches of the $g'$ family are natural, as their orbit shapes are the same except for 
    a simple rotation of 180$^{\circ}$ about the $\eta$ axis. Once $\beta>0$, the symmetry of the potential is broken and the 
    equivalent solutions are no longer as similar in shape. Indeed, it is now the two solutions of the $G'$ branch that retain similar
    shape (initially, since one is unstable).  Nevertheless, the existence of two stable equilibria and one unstable equilibrium remains
    even with finite $\beta$, as demonstrated by \cite{HK96}.
    
\begin{figure}
    \centering
\includegraphics[width=1.0\linewidth]{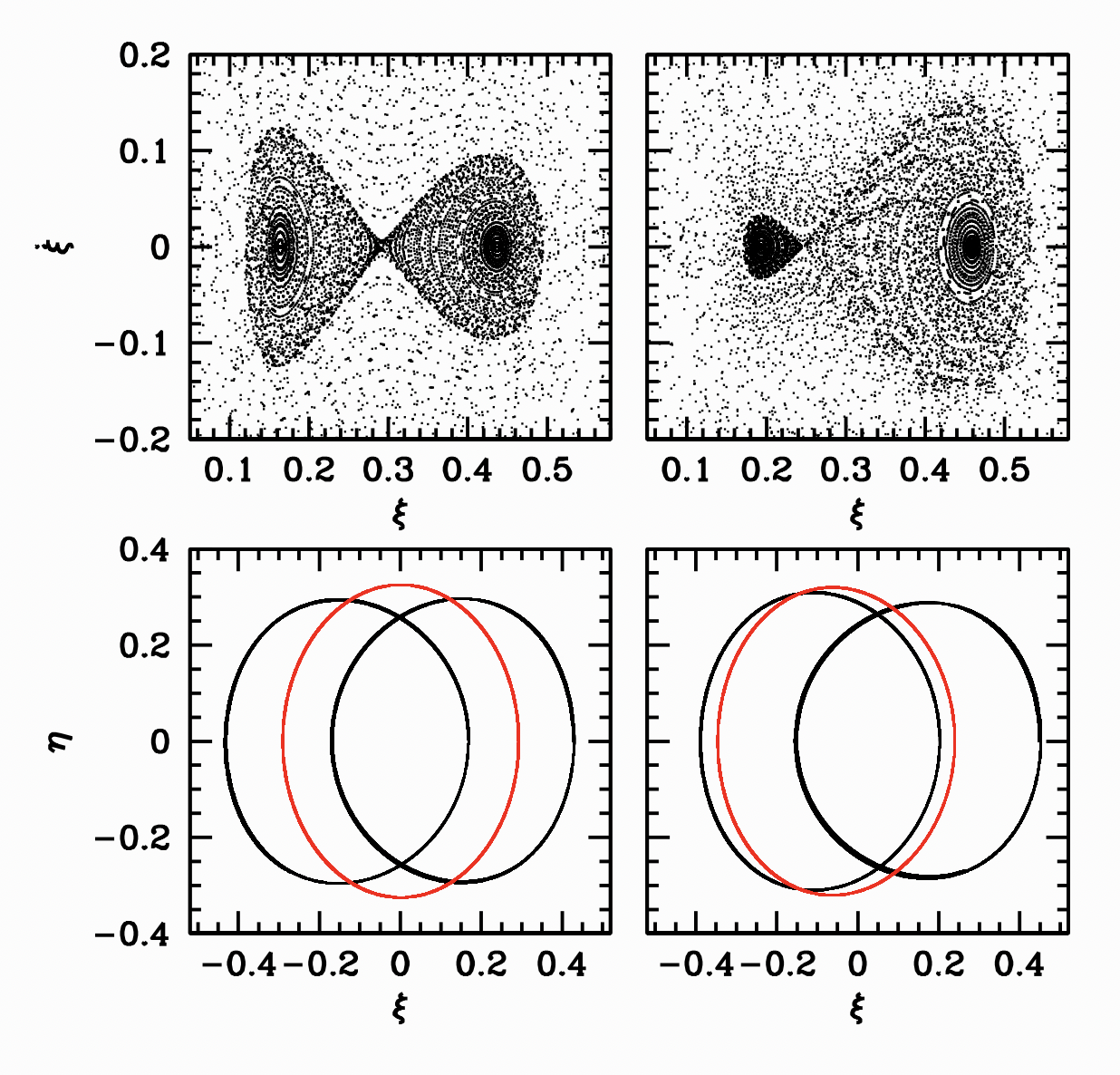}
\caption{The two panels on the left are for the case of $\beta=0$ and the two panels on the right are for
$\beta=10^{-3}$. All integrations are for the case $C_H=4.4$. In the upper panels we show the Poincare
plots. Each shows two clear stable equilibria and one saddle point between them. However, the $\beta=0$
case is much more symmetric. The differences can also be seen in the lower panels, which show the
orbits corresponding to the equilibria (red is unstable in the long term). On the left, the two stable equilibria are
close to reflection-symmetric in $\xi$, but the plot on the right shows that this symmetry has been broken
by the presence of radiation pressure. It is now the two $G'$ solutions that share a similar shape -- because they emerge together as $C_H$ drops below the critical value. \label{Example1}}
\end{figure}

      Figure~\ref{Example1} demonstrates how the orbital shapes are affected by non-zero $\beta$. The two upper panels show Poincare plots of orbits crossing the $\eta=0$ plane with $\dot{\eta}>0$,  for $C_H=4.4$ in both the $\beta=0$ and $\beta=10^{-3}$ cases.
     We see that the stable orbits are centered around the two  equilibria, tightly confined within the separatrix that passes through the unstable saddle point. The two lower panels show the shapes of thie corresponding three equilibria in the $\xi$--$\eta$ plane, with the unstable orbit shown in red. The similarity of the two $G'$ orbits to each other, for finite $\beta$, is clear.

\begin{figure}
    \centering
\includegraphics[width=1.0\linewidth]{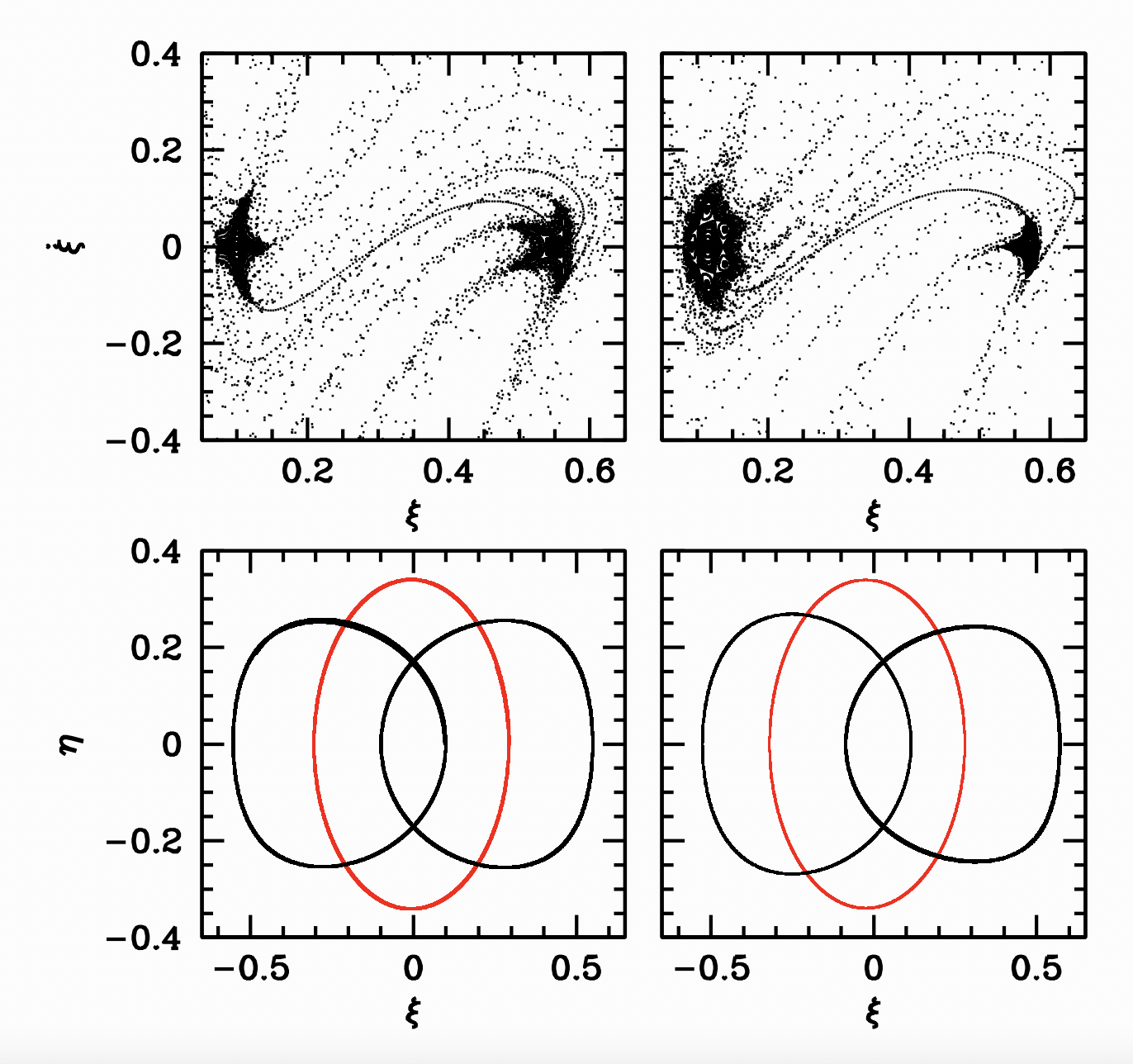}
\caption{The two panels on the left are for the case of $\beta=0$ and the two panels on the right are for
$\beta=10^{-3}$. All integrations are for the case $C_H=4.3$. In the upper panels we show the Poincare
plots. Each again shows two clear stable equilibria and one saddle point between them. The lower panels
show the orbits corresponding to the equilibria in each case. The comparison between the two lower panels
shows that the shapes of the $G$ and $G'$ orbits evolve towards similar shapes as $C_H$ drops, approaching
a similar structure as the original $\beta=0$ case.
 \label{Example2}}
\end{figure}

The $G$ family solution continues to evolve to ever more eccentric shapes as $C_H$ decreases. As $C_H$ decreases, the shape of the inner stable solution on the $G'$ branch also evolves to a more eccentric
    shape and approaches the same dichotomy with the $G$ solution as in the $\beta=0$ case. Figure~\ref{Example2} shows
    the case for $C_H=4.3$.  We see that the two equilibria are now islands of stability with an intervening unstable region, and that the remaining stable equilibria are now more similar (and more eccentric). 

As $\beta$ increases, the gap in $C_H$ between the $G$ and $G'$  families increases, and the $G'$ family eventually
becomes completely unstable for $\beta>0.01$. For $\beta>0.095$, the $G'$ family of equilibria disappears.

\subsection{Retrograde Orbits}

\begin{figure}
    \centering
\includegraphics[width=1.0\linewidth]{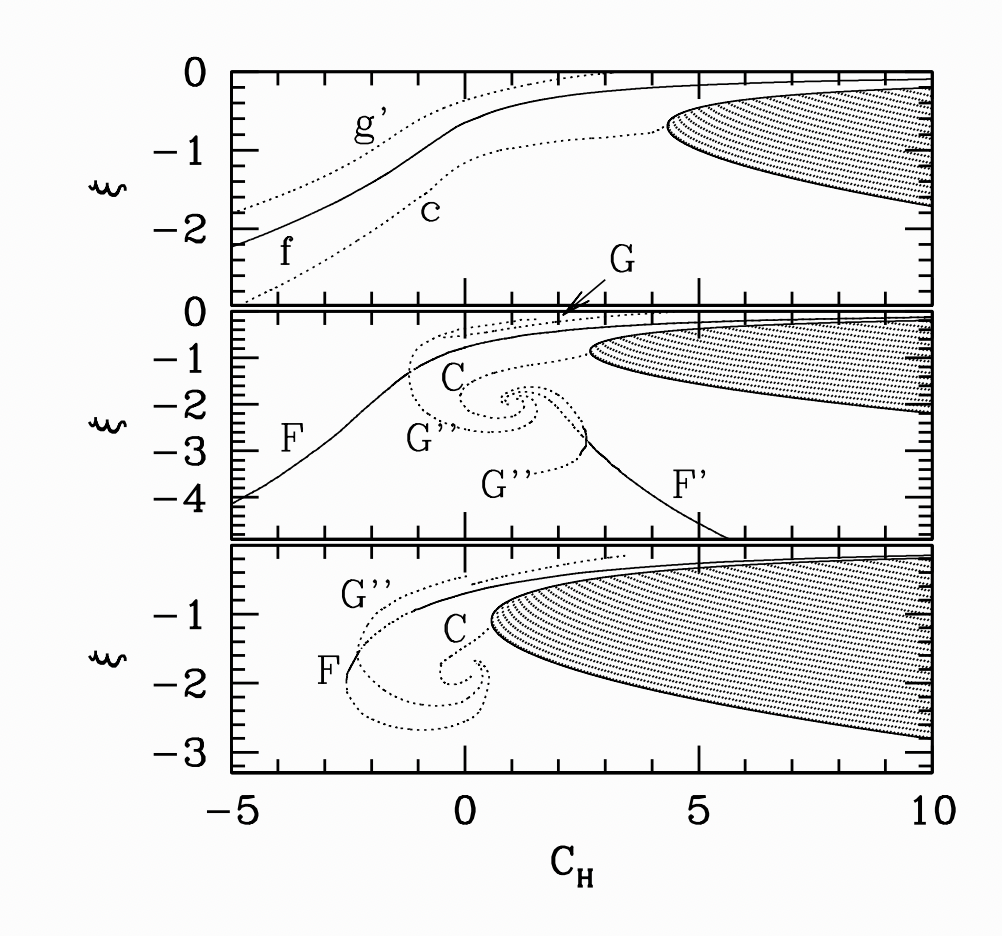}
\caption{The upper panel shows the retrograde orbital families in the original $\beta=0$ case. The
black solid curve shows the original stable family  and the black
dotted curves shows the unstable equilibrium family $c$ and the negative intersection of the $g'$ family. 
The middle panel shows the orbital families for the case $\beta=0.1$. The family $F$ is the equivalent
of $f$ and is also stable except for a brief crossing with the unstable $G''$ family. We also find the
family $F'$, associated with the second asymptotic solution, and which is also stable at large enough distances.
The shapes of the unstable families $C$,$G$ and $G''$ are affected by the presence of the triangular
equilibria, as discussed in the appendix. The bottom panel shows the case for $\beta=0.2$. We see
here that the family $F$ no longer continues to arbitary distances.
In all panels, the shaded region represents the forbidden region for that particular $\beta$.  \label{Retro}}
\end{figure}

A striking feature of Henon's original analysis was that the family $f$ of retrograde simply periodic orbits remained
stable to arbitrarily large distances from the planet, merging into the class of orbits referred to as quasi-satellites.
Figure~\ref{Retro} shows how this family adjusts to the strength of the radiation pressure.

For small values of $\beta$ the nature of the retrograde family (called $F$ here) remains similar to the $\beta=0$ case. However,
for $\beta = 0.1$, the curve shifts to larger separations more quickly, and begins to exhibit qualitatively new 
features for $\beta=0.2$. In the $\beta=0.2 $ case, the $F$ family does not extend all the way to arbitrarily
small $C_H$ but loops back to larger $C_H$ at larger $\left| \xi \right|$, forming an unstable branch. Figure~\ref{FamilyF}
in Appendix~\ref{Gallery} demonstrates this evolution.

\begin{figure}
    \centering
\includegraphics[width=1.0\linewidth]{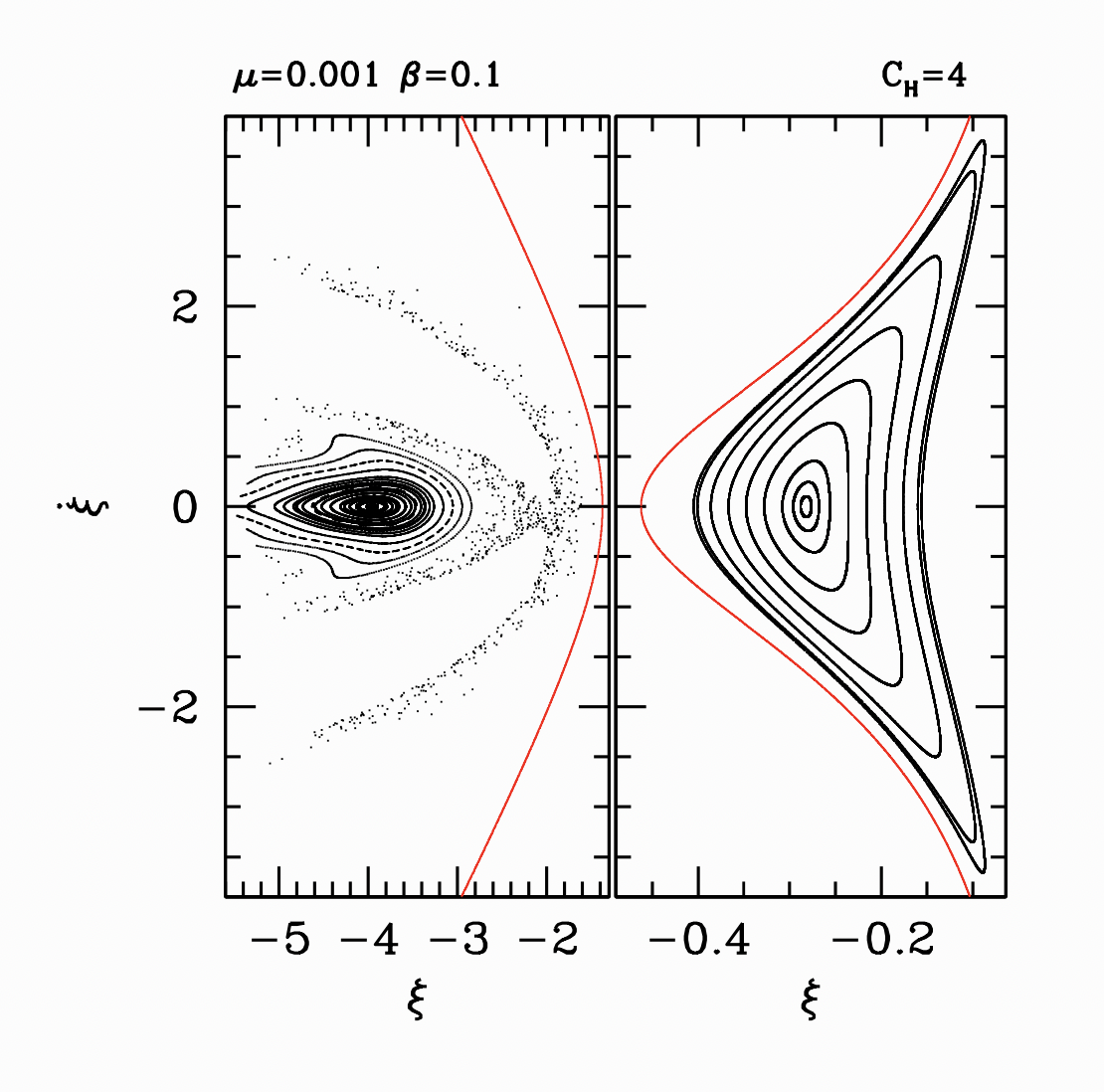}
\caption{The two panels show the Poincare plot for the retrograde orbits in the case $\mu=10^{-3}$, $\beta=0.1$
and $C_H=4$. The red curves delineate the edges of the forbidden region, which separates the $F$ region of
stability (on the right) from the $F'$ region (on the left).  In the left panel, the diffuse band of points near the red contour correspond to the quasi-circular orbits about the primary in the full restricted three-body problem. This is a diffuse band because these do not correspond to a fixed point family in the Hill problem. \label{2Sols}}
\end{figure}

The middle panel of Figure~\ref{Retro} also shows a  second
solution branch, called $F'$, which  does not appear in the $\beta$=0 case.  Appendix~\ref{AsySol}
demonstrates that this can be understood to be the result of the fact that the asymptotic solution
at large distances from the planet now splits into two branches for $\beta>0$. The solution branch $F'$ 
 corresponds to the
second  branch of this asymptotic solution. At large distances, this forms a second stable branch,
shown as the blue curves in Figure~\ref{FamilyF} in Appendix~\ref{Gallery}. The co-existence of these two stable
branches, $F$ and $F'$ is shown in Figure~\ref{2Sols} for the value $C_H=4$ and $\beta=0.1$.

Another striking feature of these equilibria is that there appears a `vortex-like' feature around $C_H \sim 1$
and $\xi \sim -1.9$, where several different equilibrium curves converge. As discussed in Appendix~\ref{Gallery},
and Figure~\ref{FamilyCon},
this appears to be the result of the interaction of the orbits with the triangular equilibrium points.  

We also show the evolution of the $G$ orbits (equivalent to the $g'$ from \cite{Hen69}). These 
result from the part of the $G$ family that becomes doubly periodic, where the condition that the path cross the $\eta=0$ axis in
the positive direction is satisfied for both positive and negative $\xi$. Thus, these curves form a
pair with curves in the prograde case (Figure~\ref{Orbit1}). 
In addition, we find another family of orbits, dubbed $G''$, which is also doubly periodic, but 
both crossings of the $\eta=0$ axis occur at $\xi<0$. These orbital families are outlined in
Figure~\ref{FamilyG} and Figure~\ref{FamilyGpp} respectively.

Finally, we note that the $F$ and $F'$ families are the only ones that extend to large distances
for $\beta=0.1$, namely that $C$ and the $G$, $G''$ families have only a finite extent. This is
consistent with the absence of any corresponding asymptotic solutions (\S~\ref{CC}).

\subsection{Available  Stable orbits}

Our goal here is to determine the parameter space available for stable orbits under the influence of both planetary
gravity 
and stellar radiation pressure. The interest in the simply periodic orbits is that stable regions are surround
these orbits, so that they form the `scaffolding' around which the structure of long-term stable orbits is built.
 Figures~\ref{Example1}, \ref{Example2} and \ref{2Sols} all demonstrate that the long-term stable orbits represent families that surround the different equilibria.

\begin{figure}
\centering
\includegraphics[width=1.0\linewidth]{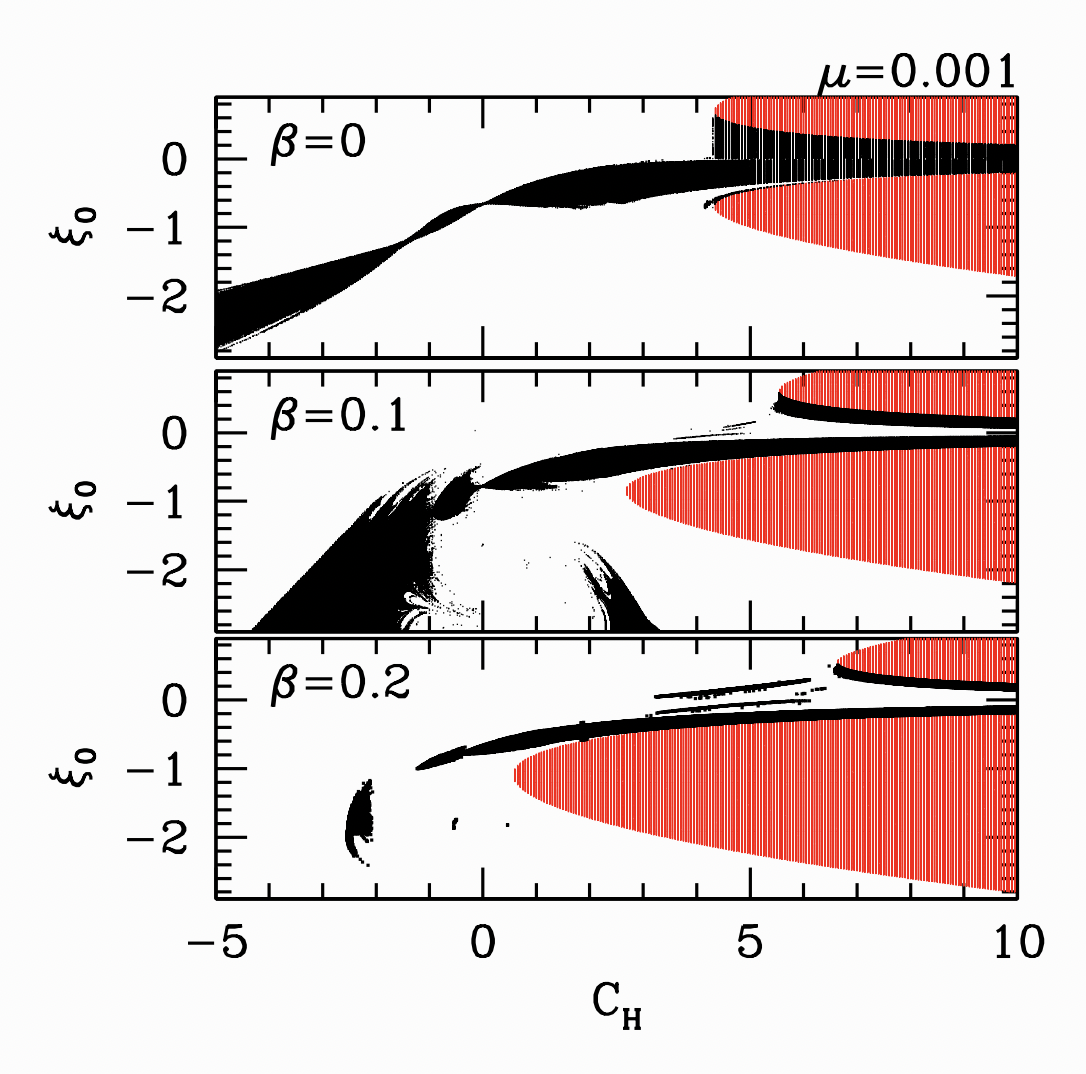}
\caption{In each panel the red region is the  range of forbidden orbits.
 Black points indicate initial conditions that produce orbits which remain within $\Delta=10$ after $t=200$ and so are
 considered stable. The strength of the radiation pressure increases from the top ($\beta=0$) to the
 botton ($\beta=0.2$). We see that the radiation pressure initially increases the range of stable distance orbits (see middle panel) but  eventually truncates the stable branch of retrograde
 orbits (lower panel). \label{Stable}}
\end{figure}

 To properly explore this parameter space,  we integrate the equations of motion for a range of initial
 conditions specified by a choice of $\xi_0$ and $C_H$, with
$\eta=0$ and $\dot{\xi}=0$, for a time=200  (where the period of the planetary orbit is 2$\pi$).
 Figure~\ref{Stable} shows the outcomes of this parameter scan
for $\mu=0.001$ and three choices of $\beta=0$, $\beta=0.1$ and $\beta=0.2$.  In this figure,
a point is plotted at the corresponding value of  $\xi_0$ and $C_H$ if the trajectory remains
within $\Delta=10$ of the planet at the end of the simulation.

The $\beta=0$ case closely resembles Figure~12 of Henon (1970), demonstrating the existence of stable orbits, both
prograde and retrograde, for $C_H>4.27$, and an extension of the stable retrograde family to arbitrarily large distances
(this family is characterised as the quasisatellites). The middle panel of Figure~\ref{Stable} shows the case of
moderate radiation pressure ($\beta=0.1$). We see that the prograde and retrograde stable regions at larger $C_H$
are now separated by a region where the orbits are unstable. This is mostly the consequence of orbits getting excited
to essentially radial orbits and hitting the planet \citep{Burns79,HK96,Zot15}.
The retrograde family still extends to larger radii and actually broadens, along with the presence of a second family
of stable orbits at larger radii and positive $C_H$. This represents the second family of asymptotic solutions, $F'$, 
 discussed in Appendix~\ref{AsySol}. 
Much of the other structure observed in Figure~\ref{Retro} does not appear in these figures because those additional
equilibria are unstable. However, the $F'$ family is robust and also appears in the full restricted three-body problem
with radiation (as can be seen from Figure~10 of \cite{Zot15}, for example -- albeit for the $\mu=0.5$ case). 

Finally, the lower panel shows the case of strong radiation pressure $\beta=0.2$. Inside the Hill sphere
the prograde and retrograde stable regions are again separated by a large region of mixed stable and unstable orbits. The retrograde family also does not now extend to arbitrarily large distances, as expected based on our asymptotic solutions. 

\subsection{Mass Dependance}

Our discussion thus far has focussed on the case of a mass ratio $\mu=10^{-3}$, appropriate for a Jupiter-mass planet
orbiting a Solar mass star. However, as we noted in our original derivation, our desire to retain the effects of non-infinitesimal
$\beta$ means that our equations are not scale free as in the case of the original Hill problem. Thus, we must investigate the effect of mass. 

\begin{figure}
\centering
\includegraphics[width=1.0\linewidth]{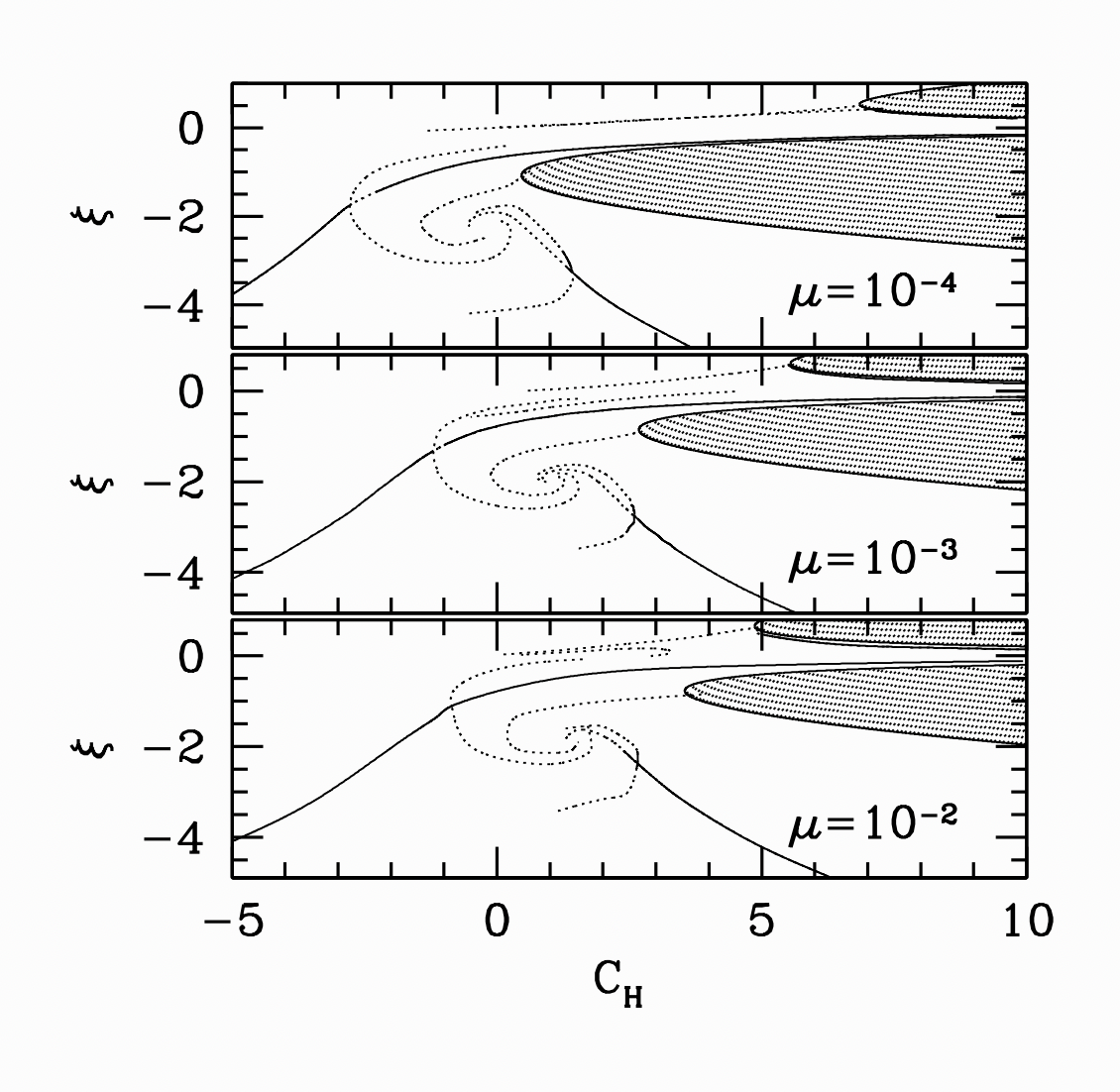}
\caption{All three panels show the orbital families for the case $\beta=0.1$. In the upper panel, the secondary is only
$\mu=10^{-4}$ of the total system mass. In the middle panel, $\mu=10^{-3}$ (so this is equivalent to the middle
panel of Figure~\ref{Retro}). In the lower panel the mass ratio is $\mu=0.01$.  Once again, stable equilibria are maked
by solid lines and unstable equilibria are marked by dotted lines. \label{Masses}}
\end{figure}

Fortunately, most of the structure in the solutions is determined by the $\beta$ parameter itself, and the role of $\mu$ is
primarily to determine a shift along the $\xi$ axis, through the effect of the $Q'$ parameter. This is demonstrated in Figure~\ref{Masses}, which shows the orbit families for the case of $\beta=0.1$ and three different mass ratios
($\mu = 10^{-4}$, $\mu=10^{-3}$ and $\mu=10^{-2}$). We see that most of the structure is maintained, but that the
lower mass families are shifted to lower $C_H$. If we examine the asymptotic solutions, we note that the frequencies
$\omega$ depend only on $\beta$, not $Q'$, and so the existence of the asymptotes should be independent of $\mu$.

 At the higher mass ratio, there are some differences in the details, as the $G'$ prograde family returns, even for $\beta=0.1$.

\section{The Full Restricted Three-Body Problem and Observational Signatures}
\label{Obs}

The direct observation of a self-luminous exoplanet requires the detection of an unresolved point source, usually at
the infrared wavelengths, where self-powered emission is expected to peak. However, if there is a substantial population
of circumplanetary dust, the scattering of stellar light by the dust may provide an observable signature at shorter
wavelengths more characteristic of the stellar emission. Similarly, with a large enough surface area, the thermal
emission from the dust may overwhelm that of the planet, although this is a function of wavelength \citep{KW11}.
Furthermore, such populations are more extended and thus potentially resolvable.  Therefore, we wish to now
place our results of the previous section within the context of the full restricted three body problem, and to identify
potentially observable signatures. We will focus primarily here on the stable orbits, which could be populated
by long-lived dust populations, and on those families within or near the planetary Hill sphere. There will
also be long-lived families associated with the $L_4$ and $L_5$ points in the non-zero $\beta$ case, but we will not
treat them here.

\subsection{The shape of a trapped dust population}

\begin{figure}
\centering
\includegraphics[width=1.0\linewidth]{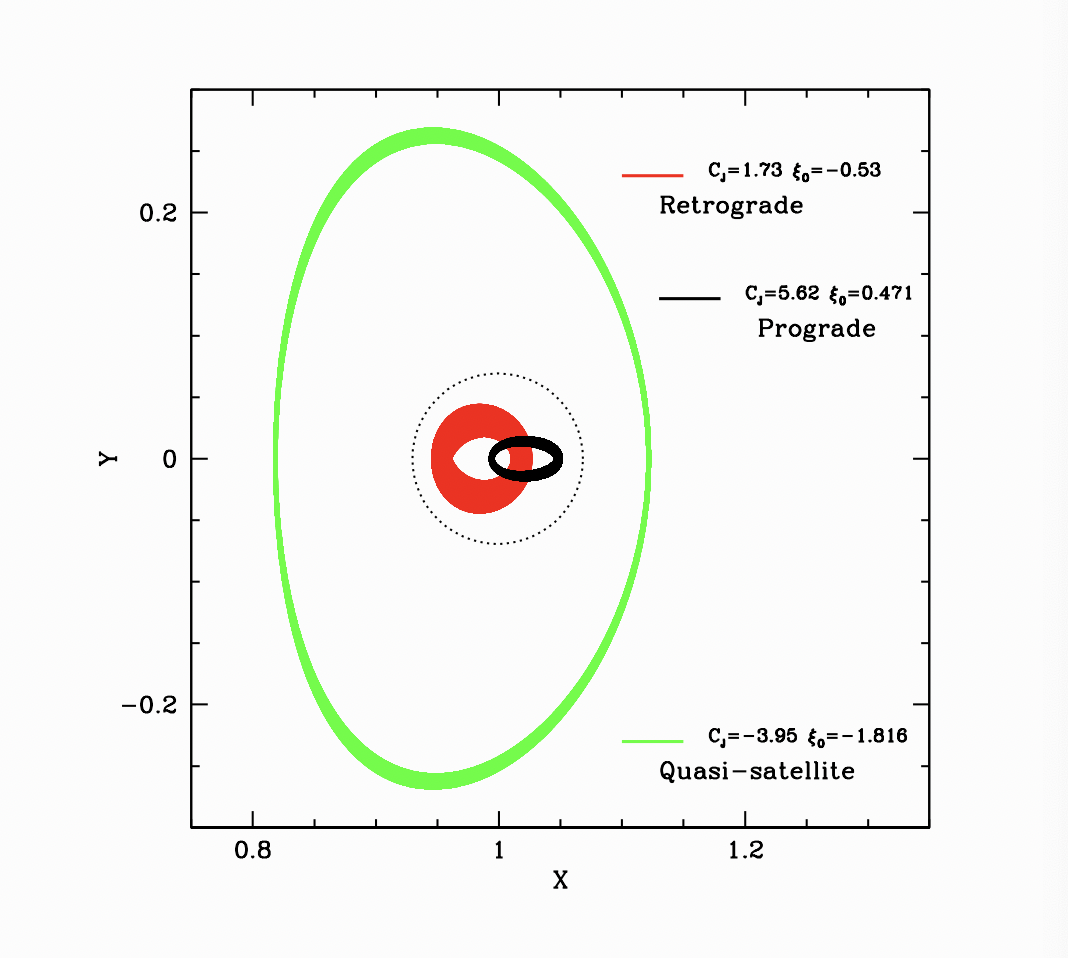}
\caption{The black orbit represents one of the widest prograde orbits that is stable, in the case of $\mu=10^{-3}$ and $\beta=0.1$. The red curve represents a stable retrograde orbit and the green curve represents a retrograde quasi-satellite orbit. The dotted curve represents the Hill sphere radius, centered on the secondary at (0.999,0). All orbits are shown in the reference frame co-rotating with the secondary. }
 \label{3bEx}
\end{figure}

To model the potential signature of a circumplanetary dust population,
 we repeat the orbital integrations of the prior sections, but now within the context of the full restricted three-body
problem, so that $X=1 - \mu + \mu^{1/3} \xi$ and $Y = \mu^{1/3} \eta$. Figure~\ref{3bEx} shows three examples of
the resulting orbits, chosen to represent the different classes of bound orbit, in the case $\mu=10^{-3}$ and $\beta=0.1$. In each case, the choice of initial parameters
was chosen from the middle of the corresponding stability region in Figure~\ref{Stable}, albeit integrated now in the full
restricted three body potential. So, these do not correspond exactly to the relevant simply periodic orbits, but are representative of the broader family of stable orbits that precess about the periodic orbit. The three examples shown are for one of the widest stable prograde orbits (black curve), a retrograde orbit of similar width (red curve) and a quasi-satellite retrograde orbit (green curve). 

 To properly represent the observations, we must go beyond the two dimensional orbits. Thus 
 we now evolve the full three
dimensional   set of equations, and choose our initial particle trajectories with positions 
  initially drawn from a singular isothermal sphere
radius distribution, spherically distributed except that we exclude starting positions with inclinations between
$60^{\circ}$--$130^{\circ}$ of the orbital plane. This is based on the assumption that the dust initial distribution should
follow that of the parent bodies. The observed irregular satellites in the Solar system show this deficit \citep{JH07}, which
is believed to be a consequence of the dynamical instability of such orbits \citep{HB91,NADL03}.  We choose initial velocities
based on the circular velocity about the planet and random orientations. We integrate these forward for $t=200$ in our
time units (where $t=2 \pi$ is one orbital period for the secondary about the primary).

\begin{figure}
\centering
\includegraphics[width=1.0\linewidth]{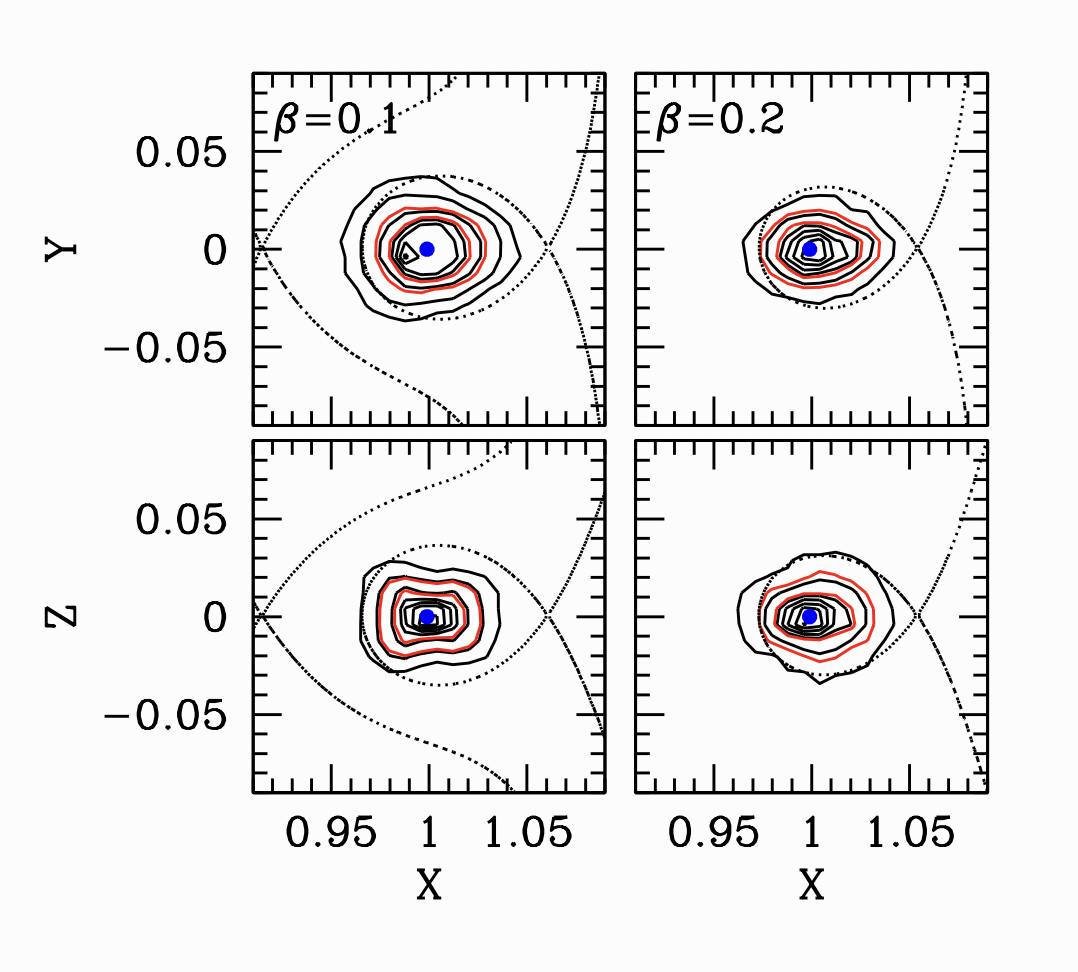}
\caption{The contours represent surface density of points for particles orbiting about a secondary with mass
ratio $\mu=10^{-3}$ and radiation pressure $\beta=0.1$ (left panels) and $\beta=0.2$ (right panels). The blue solid point represents the position of the secondary. The red solid contours represent the value corresponding to half the maximum and a value a quarter of the maximum. The dotted contours represents the equipotential that pass through the $L_1$ and  $L_2$ points.   It is the confinement within the $L_2$ contour that is responsible for the elongated shape of the distribution. The upper panel represents the X-Y plane, and the lower panel the X-Z plane.\label{Shape}}
\end{figure}

Figure~\ref{Shape} shows contour plots of the resulting density of particles for the cases $\beta=0.1$ (left panels) and
 $\beta=0.2$ (right panels), in both
the X-Y plane (at Z=0) and in the X-Z plane (at Y=0). The distribution of particles
is primarily confined within the seperatrices that pass through $L_2$ and it is this surface which is responsible for the
elongation of the observed distribution. The half-max and quarter-max contours are marked in red, demonstrating this
elongation, in both the X-Y and X-Z planes.  The principal feature is the elongation of the observable population along
the axis between the secondary and the primary. This is to be expected given the ellipticity of the orbits shown
in Figure~\ref{3bEx}, and also discussed in \cite{HK96}. In particular, we see that the combination of prograde and retrograde orbits are expected to give a tapered appearance to the contours, with the flattening increasing away from the direction of the primary.

If we characterise the shape of the original parent population, by following the above model with $\beta=0$, we find
that the Full-Width Half Max (FWHM) of the surviving particles is $0.28$ of the Hill diameter in the X direction, $0.27$ in the $Y$ direction,
and 0.20 in the $Z$ direction. These proportions increase to 0.39, 0.37 and 0.30 respectively, if we take the widths at
quarter of the maximum (FWQM). Thus, the underlying population is spherically symmetric in projection, despite the Kozai-induced
holes at the pole.

The results of the $\beta=0.1$ integration show FWHM of 0.30 in the $X$ direction, 0.24 in the $Y$ direction and $0.20$ in
the $Z$ direction (with 0.40, 0.31 and 0.29 at FWQM). Thus, the distribution of surviving dust is  elongated with an aspect
ratio $\sim 1.3:1$. For $\beta=0.2$ the  elongation becomes even more extreme, with FWHM of $0.31$ in $X$, 0.21 in $Y$
and 0.17 in $Z$ (with 0.43, 0.29 and 0.33 respectively at FWQM). Thus, the elongation of the dust population
increases with increasing $\beta$.

This suggests a potential signature  of the influence of radiation pressure -- the observation of an extended distribution of dust yields a potentially resolvable target in
thermal or scattered emission, and one that should become more elongated as one approaches shorter wavelengths that probe smaller particles and larger $\beta$.

\subsection{Fomalhaut: A case study}

Although the above model can be applied to any star/planet system imaged in scattered or thermal light, the
 star Fomalhaut and its attendant dust structures offers the most complete application to date. The narrow dust
 ring imaged by \cite{Kalas05} has long stimulated discussion about the origin of this dust and how planets
 might sculpt it \citep{Q06,CKK09}, and the detection of a candidate planet  Fomalhaut~b \citep{Kalas08,Kalas13} has
 further amplified this discussion, including alternative interpretations \citep{Currie12}. 
 
 The narrowness of the dust ring is another feature that needs explanation. \cite{HH23} offer an alternative
 to the standard birth ring of planetesimals --possibly sculpted by additional planets \citep{BPC12}, in which
 the dust is generated in a circumplanetary cloud of irregular satellites and then spirals out through the $L_2$
 point, sherpherded into a thin ring by the restrictions imposed by the relevant Jacobi constant.
 The discussion of \cite{HH23} focusses on the ring morphology, but the potential observability of the dust bound
 to the planet is a further interesting consideration. \cite{KW11} discuss the application of this model to the
 observations of Fomalhaut~b, but can be considered relevant to any putative irregular satellite cloud, either at the
 location of Fomalhaut~b or elsewhere in the system. 
 
 \cite{KW11} discuss a dust cloud whose spatial distribution follows that of the parent bodies, but our calculations
 above show that the dust dynamics under the influence of radiation pressure result in an elongated structure, with
 the long axis pointing along the star--planet line. Interestingly, such a signature was discussed by \cite{Kalas13}
 for Fomalhaut~b, although it could not be discounted that this was a residual signature of speckle noise. 
 
 An additional signature of the \cite{HH23} model is that the planet should lie slightly interior to the inner edge
 of the dust ring, by virtue of the geometry of the equipotential that passes through the $L_2$ point (through which
 the unbound dust escapes). This implies that the planet should lie at $\sim 0.92$ the semi-major axis as measured
 by the shape of the ring. Once again, Fomalhaut~b lies at about the correct distance from the ring, although the
 proper motion of the planet does not conform to this model any more than it does for the gravitational sculpting
 model of \cite{Q06,CKK09}. 
 
 
 
 \subsubsection{Fom b as a quasi-satellite?}
 \label{QuasiFom}.
 
 The fading and eventual disappearance of Fomalhaut~b \citep{Kalas13,GR20} in the optical suggests that this object
 is not a planet but a dust cloud undergoing dispersal over a timescale of decades. The origin of the dust cloud has been postulated
 to be the result of a collision between planetesimals \citep{Currie12,Kalas13,GR20} although the origin of the parent
 population is still somewhat in question (since it is not located in the nominal birth ring associated with the dust disk).
 The discovery of interior belts \citep{GWR23} is a possible source, although the path from the quasi-circular interior belts to a
 coherent, highly elliptical, dispersing orbit has yet to be described in detail.
 
 The population of quasi-satellites associated with Henon's $f$ family offers a potential parent population. If the planet
 associated with the dust disk is accompanied by a population of planetesimals in quasi-satellite orbits, then these
 objects would have heliocentric orbits with semi-major axes within $\sim 20\%$ of the planetary orbit, and with
 eccentricities $\sim 0.1$--$0.3$. However, if such an object were to release dust that is subject to radiation pressure, the
 orbit of the dust would deviate from that of the parent body, because the $F$ and $F'$ orbits in Figures~\ref{Retro} and \ref{Stable}
 are not the same as for the $\beta=0$ case. Indeed, the inferred heliocentric parameters for particles on these
 orbits can deviate substantially from those characteristic of the planet itself, even reaching apparently unbound values
 for part of the orbit (see appendix~\ref{HelioFp} for an example).
 
 \begin{figure}
\centering
\includegraphics[width=1.0\linewidth]{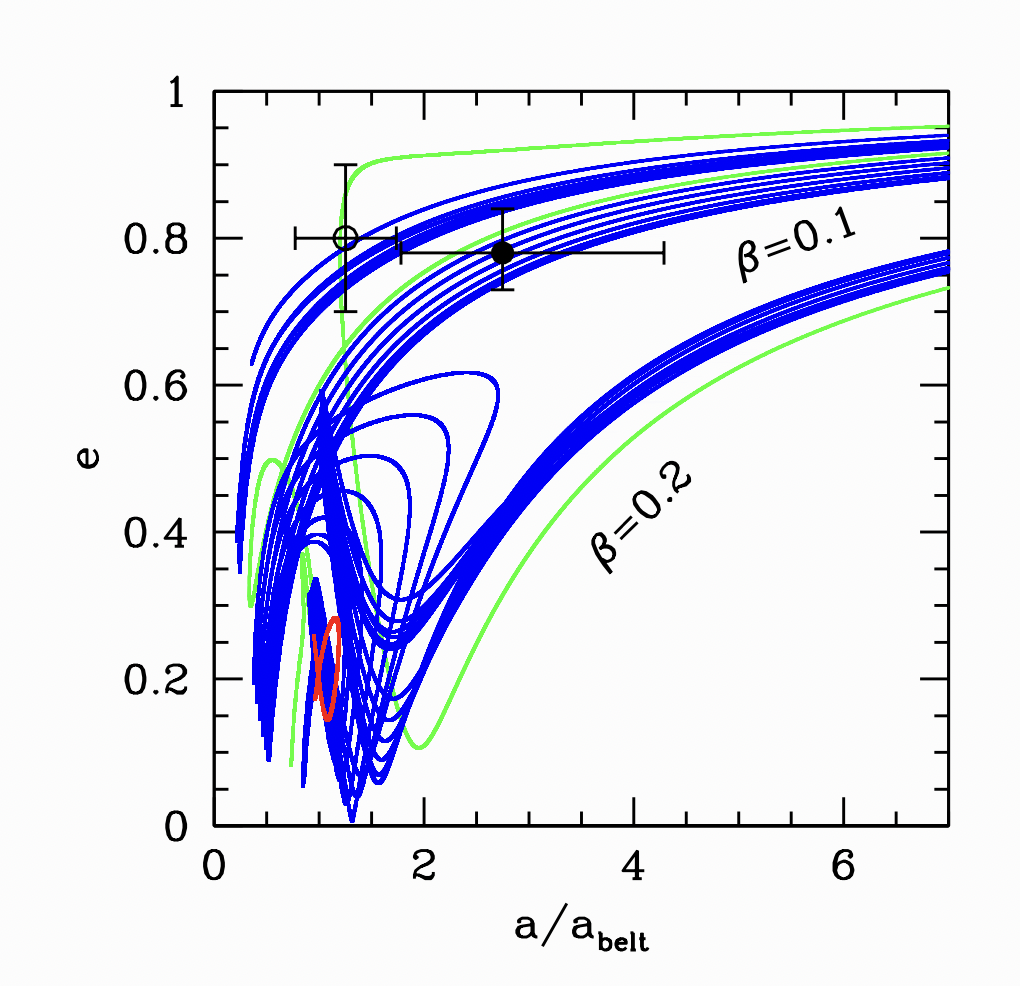}
\caption{The solid point shows the estimated orbital parameters of Fom~b from
Gaspar \& Rieke (2020), while the open point represents the earlier measurement of Kalas et al. (2013). 
The red curve shows the evolution of the heliocentric parameters of quasi-satellite of a Jupiter-mass planet orbiting at the distance of the primary Fomalhaut dust belt. The blue curve shows the orbital evolution if we start with the same initial conditions as the red curve, but assume radiation pressure with $\beta=0.1$. The green curve corresponds to $\beta=0.2$.  Semi-major axes are normalised by
the observed semi-major axis of the debris belt, with the planet offset relative to that according to the model
of \cite{HH23}. \label{QuasiB}}
\end{figure}

Figure~\ref{QuasiB} demonstrates how radiation pressure would cause the dust to deviate.
 Three orbits are shown. The red trajectory shows the
variation in semi-major axis (scaled relative to the parent planet) and eccentricity of a  single quasi-satellite trajectory in the
case $\beta=0$. If one takes the same initial position and velocity, but calculates the trajectory with $\beta=0.1$ (blue curve)
and $\beta=0.2$ (green curve), one finds that the dust   orbit makes several loops
in semi-major axis and eccentricity, which pass through the region occupied by the measured orbital parameters of
Fomalhaut~b. For $\beta=0.2$, the dust does pass through the observed region of parameter space, but is ejected
pretty quickly. However, for $\beta=0.1$, the orbit lies within the stable part associated with the $F'$ family, and 
makes repeated passages through the observed region. These results suggest that the dust involved in the cloud should have $\beta \sim 0.1$ to increase the chances of observation.
 For values
close to $\beta \sim 0$, the orbital parameters do not approach those observed, and for $\beta \sim 0.2$, the dust leaves
the system rapidly.   For $\beta \sim 0.1$, orbits of this type complete many orbits around the primary, spending $\sim 4\%$ of their time within the
inferred eccentricity range. This is consistent with the fact that the observations from 2008--2020 span $<1\%$ of the estimated
orbital period at the inferred separation of Fom~b, and can be described by a single set of osculating orbital parameters.

If Fom~b is a dissipating dust cloud, the rate at which such objects are generated is very uncertain.  It takes $\sim 700$ years
for the eccentricity to grow to the observed value from the original $\beta=0$ orbit, so that any individual event has a low
probability of detection within a 10 year window.
 The fact that one has
been observed over the course of our brief observations suggest that the rate  of dust cloud generation is non-negligible (or we have been very
lucky). Although it may be challenging to justify a high rate of collision in the system, we should note that the Solar system contains populations of `active asteroids' that need not experience collisions to eject dust \cite[e.g.][]{ActiveA}. 
If more such dust clouds are discovered in the future, their kinematics should be similar to that described in Figure~\ref{QuasiB} in this model, whereas they should exhibit a much wider range of parameters if they are being seeded
by the interior belts observed in the system.

 \subsubsection{Fom b as an $L_6$ point?}

Much has been made of the curious orbit that Fom~b appears to trace. Most of the attempted physical explanations assume an orbit close to coplanar with the disk. If we were to hypothesize that this feature was, instead, associated with the out-of-plane equilibria due to radiation pressure, could this very different geometry provide a better explanation for Fom~b?
The short answer is no. 
 One can plot  the locus of possible positions of the $L_6$ and $L_7$ points, assuming that the Fomalhaut ring
 indicates the orbital plane for any secondary.  For different values of $\beta$, these points lie at different heights above the plane which, in turn, project to different locations
on the sky plane. However, none of them come close to Fomalhaut~b. 

Somewhat more strikingly, the resulting locus  does pass close to the location of the feature dubbed the `Great Dust Cloud' in the
JWST images of this system \citep{GWR23}. However, it requires very small particles to achieve $\beta>1$, and so the
23.5~$\mu$m bandpass seems like an unlikely wavelength to identify such features. Furthermore, ground-based imaging suggests this feature may be a background feature \citep{KLKF} and so this agreement appears to be a coincidence.


\subsubsection{Searching for the True Planet}

We have thus far focussed on the potential shape of a population of trapped dust. To estimate the brightness, we
must account for the finite lifetime of dust trapped within the Hill sphere, subject to both erosion due to collisional
evolution, and to the loss of material through the $L_2$ point to the thin disk. In \cite{HH23} we estimated the
average time for escape to be $\sim 10^7$~years for grains with $\beta \sim$ 0.1--0.2 in orbit around a Jupiter mass planet at 140~AU.  In similar fashion as \cite{KW11}, we can estimate the timescale in a population of irregular satellites  to generate dust
by collisions. In a collisional cascade, the mass and  mass loss rate is dominated by the most massive bodies, 
 denoted by radius $s_{max}$, but the surface area
 for emission and scattering is dominated by the smaller bodies,  of radius s.
 The rate of collisions amongst the larger bodies is 
\begin{eqnarray}
T_{coll} & \sim&  4 \, {\rm Myr} \left( \frac{a}{140 AU} \right)^{\frac{7}{2}}  \\ 
&& \left(\frac{s}{10 \mu m} \right) \left(\frac{s_{max}}{100 km}\right) 
\left( \frac{M_p}{1 M_J} \right)^{\frac{2}{3}} \left( \frac{M_{irr}}{M_{\leftmoon}} \right)^{-1} \nonumber
\end{eqnarray}
assuming a total mass  $M_{irr}$,  scaled here relative to a Lunar mass, confined to a volume $\sim 0.3$ Hill radii. We truncate the collisional
cascade at dust radii $s \sim 10 \mu$m, because $\beta \sim 0.18 (10 \mu m/s)$ for the Fomalhaut system and this
is approximately the $\beta$ value where dust escapes before being ground down further \citep{HH23}.  Thus,
an irregular satellite cloud of this size would be in approximate steady state. We note that this age is younger than
the overall age of the Fomalhaut system, but we expect irregular satellites to be captured during three-body interactions
during epochs of dynamical instability in a planetary system, which can occur at ages of hundreds of millions of years (as has
 been hypothesized in our own Solar system).
  Assuming a \cite{Dohn69} power law,
 extending down from 100 km to 10$\mu$m, we estimate a total  cross-sectional area  in dust $\sim 1.5 \times 10^{23} cm^2$ ( similar to that estimated by \cite{KW11} for this system with slightly different assumptions). 
 
 To assess whether this is observable in the recent JWST images \citep{GWR23}, we estimate that the dust should
 be at a temperature $\sim 47 {\rm K} \left( a/140 AU \right)^{-1/2}$.  At 23$\mu$m, and at a distance of 7.7~pc, this
 implies an object with total observable flux $\sim 0.4 \mu$Jy, or $\sim 1 \mu$Jy per square arcsecond, if spread over an angular
 width of $\sim 0.4 ''$ (corresponding to 0.3 Hill radii).  This is $\sim 150$ times fainter than the `Great Dust Cloud'
 and about ten times fainter than the $1 \sigma$ noise level estimated by \cite{GWR23}. 
 
 The above estimate assumes a blind search, but one might also help to narrow the field if one associates the Fom~b
 orbit with an associated quasi-satellite. However, 
quasi-satellites can exhibit large excursions relative to their guiding center planets, so the identification of Fom~b
with such a population does not narrow the location of any particular planet (other than locating it in the western
half of the orbit).

We conclude that the nominal model is not yet easily detectable with the extant JWST data, but does lie within an
order of magnitude of the current noise levels. Thus, either deeper observations or a more optimistic estimate for the
size of the dust population could bring observations and theory into comparable brightness levels.

\section{Conclusions}

The calculations in this paper represent a follow-on from the calculations of \cite{HH23}. In particular, if we posit that 
geometrically thin dust rings are the product of the grinding down of an irregular satellite population trapped within the
Hill sphere of an extrasolar planet, then we seek to identify observational signatures of a dust population that remains
bound to the planet.

To that end, we have studied the orbital properties of dust affected by radiation pressure and the gravity of the secondary body, and classified the kinds of periodic solutions that may occur. We have extended our study to values
of $\beta$ that exceed the infinitesimal limit treated in many previous studies. We find that the orbital families of the
$\beta=0$ case are reproduced for small $\beta$ but some of them disappear or are substantially modified at larger
values of $\beta$. 

The modified orbital properties  at finite $\beta$ suggest that a bound dust cloud should be elongated along the primary--secondary
axis, with the level of elongation increasing with $\beta$. This suggests that resolved sources whose shapes change
with the observed wavelength may be the signature of radiation pressure at work.

As an application of our model we consider the Fomalhaut system and its enigmatic planet/dust cloud Fomalhaut~b. 
We find that the curious orbital properties of Fom~b could be explained if the parent body of the dust cloud was following
a quasi-satellite trajectory relative to a yet-undiscovered planet orbiting in the narrow dust disk. The effect of radiation pressure on dust released from such bodies can produce orbital elements consistent with those observed.

{\bf Data availability}: The data underlying this article will be shared on reasonable request to the corresponding author.

The author thanks the referees for constructive comments.
 This research was supported by NASA grant  80NSSC20K0266. This research has made use of NASA's Astrophysics Data System Bibliographic Services. 
 
\bibliographystyle{mnras}
\bibliography{refs}

\begin{thebibliography}{}
\makeatletter
\relax
\def\mn@urlcharsother{\let\do\@makeother \do\$\do\&\do\#\do\^\do\_\do\%\do\~}
\def\mn@doi{\begingroup\mn@urlcharsother \@ifnextchar [ {\mn@doi@}
  {\mn@doi@[]}}
\def\mn@doi@[#1]#2{\def\@tempa{#1}\ifx\@tempa\@empty \href
  {http://dx.doi.org/#2} {doi:#2}\else \href {http://dx.doi.org/#2} {#1}\fi
  \endgroup}
\def\mn@eprint#1#2{\mn@eprint@#1:#2::\@nil}
\def\mn@eprint@arXiv#1{\href {http://arxiv.org/abs/#1} {{\tt arXiv:#1}}}
\def\mn@eprint@dblp#1{\href {http://dblp.uni-trier.de/rec/bibtex/#1.xml}
  {dblp:#1}}
\def\mn@eprint@#1:#2:#3:#4\@nil{\def\@tempa {#1}\def\@tempb {#2}\def\@tempc
  {#3}\ifx \@tempc \@empty \let \@tempc \@tempb \let \@tempb \@tempa \fi \ifx
  \@tempb \@empty \def\@tempb {arXiv}\fi \@ifundefined
  {mn@eprint@\@tempb}{\@tempb:\@tempc}{\expandafter \expandafter \csname
  mn@eprint@\@tempb\endcsname \expandafter{\@tempc}}}

\bibitem[\protect\citeauthoryear{{Boley}, {Payne}, {Corder}, {Dent}, {Ford}  \&
  {Shabram}}{{Boley} et~al.}{2012}]{BPC12}
{Boley} A.~C.,  {Payne} M.~J.,  {Corder} S.,  {Dent} W.~R.~F.,  {Ford} E.~B.,
  {Shabram} M.,  2012, \mn@doi [\apjl] {10.1088/2041-8205/750/1/L21}, \href
  {https://ui.adsabs.harvard.edu/abs/2012ApJ...750L..21B} {750, L21}

\bibitem[\protect\citeauthoryear{{Burns}, {Lamy}  \& {Soter}}{{Burns}
  et~al.}{1979}]{Burns79}
{Burns} J.~A.,  {Lamy} P.~L.,   {Soter} S.,  1979, \mn@doi [\icarus]
  {10.1016/0019-1035(79)90050-2}, \href
  {https://ui.adsabs.harvard.edu/abs/1979Icar...40....1B} {40, 1}

\bibitem[\protect\citeauthoryear{{Chiang}, {Kite}, {Kalas}, {Graham}  \&
  {Clampin}}{{Chiang} et~al.}{2009}]{CKK09}
{Chiang} E.,  {Kite} E.,  {Kalas} P.,  {Graham} J.~R.,   {Clampin} M.,  2009,
  \mn@doi [\apj] {10.1088/0004-637X/693/1/734}, \href
  {https://ui.adsabs.harvard.edu/abs/2009ApJ...693..734C} {693, 734}

\bibitem[\protect\citeauthoryear{{Currie} et~al.,}{{Currie}
  et~al.}{2012}]{Currie12}
{Currie} T.,  et~al., 2012, \mn@doi [\apjl] {10.1088/2041-8205/760/2/L32},
  \href {https://ui.adsabs.harvard.edu/abs/2012ApJ...760L..32C} {760, L32}

\bibitem[\protect\citeauthoryear{{Dohnanyi}}{{Dohnanyi}}{1969}]{Dohn69}
{Dohnanyi} J.~S.,  1969, \mn@doi [\jgr] {10.1029/JB074i010p02531}, \href
  {https://ui.adsabs.harvard.edu/abs/1969JGR....74.2531D} {74, 2531}

\bibitem[\protect\citeauthoryear{{Domingos}, {Winter}  \&
  {Yokoyama}}{{Domingos} et~al.}{2006}]{DWY06}
{Domingos} R.~C.,  {Winter} O.~C.,   {Yokoyama} T.,  2006, \mn@doi [\mnras]
  {10.1111/j.1365-2966.2006.11104.x}, \href
  {https://ui.adsabs.harvard.edu/abs/2006MNRAS.373.1227D} {373, 1227}

\bibitem[\protect\citeauthoryear{{Galicher}, {Marois}, {Zuckerman}  \&
  {Macintosh}}{{Galicher} et~al.}{2013}]{Gal13}
{Galicher} R.,  {Marois} C.,  {Zuckerman} B.,   {Macintosh} B.,  2013, \mn@doi
  [\apj] {10.1088/0004-637X/769/1/42}, \href
  {https://ui.adsabs.harvard.edu/abs/2013ApJ...769...42G} {769, 42}

\bibitem[\protect\citeauthoryear{{Garc{\'\i}a Y{\'a}rnoz}, {Scheeres}  \&
  {McInnes}}{{Garc{\'\i}a Y{\'a}rnoz} et~al.}{2015}]{YSM15}
{Garc{\'\i}a Y{\'a}rnoz} D.,  {Scheeres} D.~J.,   {McInnes} C.~R.,  2015,
  \mn@doi [Celestial Mechanics and Dynamical Astronomy]
  {10.1007/s10569-015-9604-9}, \href
  {https://ui.adsabs.harvard.edu/abs/2015CeMDA.121..365G} {121, 365}

\bibitem[\protect\citeauthoryear{{Gaspar} \& {Rieke}}{{Gaspar} \&
  {Rieke}}{2020}]{GR20}
{Gaspar} A.,  {Rieke} G.,  2020, \mn@doi [Proceedings of the National Academy
  of Science] {10.1073/pnas.1912506117}, \href
  {https://ui.adsabs.harvard.edu/abs/2020PNAS..117.9712G} {117, 9712}

\bibitem[\protect\citeauthoryear{{G{\'a}sp{\'a}r} et~al.,}{{G{\'a}sp{\'a}r}
  et~al.}{2023}]{GWR23}
{G{\'a}sp{\'a}r} A.,  et~al., 2023, \mn@doi [Nature Astronomy]
  {10.1038/s41550-023-01962-6}, \href
  {https://ui.adsabs.harvard.edu/abs/2023NatAs.tmp...93G} {}

\bibitem[\protect\citeauthoryear{{Giancotti}, {Campagnola}, {Tsuda}  \&
  {Kawaguchi}}{{Giancotti} et~al.}{2014}]{GCT14}
{Giancotti} M.,  {Campagnola} S.,  {Tsuda} Y.,   {Kawaguchi} J.,  2014, \mn@doi
  [Celestial Mechanics and Dynamical Astronomy] {10.1007/s10569-014-9564-5},
  \href {https://ui.adsabs.harvard.edu/abs/2014CeMDA.120..269G} {120, 269}

\bibitem[\protect\citeauthoryear{{Giuppone}, {Beaug{\'e}}, {Michtchenko}  \&
  {Ferraz-Mello}}{{Giuppone} et~al.}{2010}]{GBM10}
{Giuppone} C.~A.,  {Beaug{\'e}} C.,  {Michtchenko} T.~A.,   {Ferraz-Mello} S.,
  2010, \mn@doi [\mnras] {10.1111/j.1365-2966.2010.16904.x}, \href
  {https://ui.adsabs.harvard.edu/abs/2010MNRAS.407..390G} {407, 390}

\bibitem[\protect\citeauthoryear{{Hamilton} \& {Burns}}{{Hamilton} \&
  {Burns}}{1991}]{HB91}
{Hamilton} D.~P.,  {Burns} J.~A.,  1991, \mn@doi [\icarus]
  {10.1016/0019-1035(91)90039-V}, \href
  {https://ui.adsabs.harvard.edu/abs/1991Icar...92..118H} {92, 118}

\bibitem[\protect\citeauthoryear{{Hamilton} \& {Krivov}}{{Hamilton} \&
  {Krivov}}{1996}]{HK96}
{Hamilton} D.~P.,  {Krivov} A.~V.,  1996, \mn@doi [\icarus]
  {10.1006/icar.1996.0175}, \href
  {https://ui.adsabs.harvard.edu/abs/1996Icar..123..503H} {123, 503}

\bibitem[\protect\citeauthoryear{{Hayakawa} \& {Hansen}}{{Hayakawa} \&
  {Hansen}}{2023}]{HH23}
{Hayakawa} K.~T.,  {Hansen} B. M.~S.,  2023, \mn@doi [\mnras]
  {10.1093/mnras/stad1091}, \href
  {https://ui.adsabs.harvard.edu/abs/2023MNRAS.522.2115H} {522, 2115}

\bibitem[\protect\citeauthoryear{{Henon}}{{Henon}}{1969}]{Hen69}
{Henon} M.,  1969, \aap, \href
  {https://ui.adsabs.harvard.edu/abs/1969A&A.....1..223H} {1, 223}

\bibitem[\protect\citeauthoryear{{Henon}}{{Henon}}{1970}]{Hen70}
{Henon} M.,  1970, \aap, \href
  {https://ui.adsabs.harvard.edu/abs/1970A&A.....9...24H} {9, 24}

\bibitem[\protect\citeauthoryear{{Hill}}{{Hill}}{1878a}]{Hill78a}
{Hill} G.~W.,  1878a, Am. J. Math., \href {none} {1, 5}

\bibitem[\protect\citeauthoryear{{Hill}}{{Hill}}{1878b}]{Hill78b}
{Hill} G.~W.,  1878b, Am. J. Math., \href {none} {1, 129}

\bibitem[\protect\citeauthoryear{{Hill}}{{Hill}}{1878c}]{Hill78c}
{Hill} G.~W.,  1878c, Am. J. Math., \href {none} {1, 245}

\bibitem[\protect\citeauthoryear{{Hunter}}{{Hunter}}{1967}]{Hunter67}
{Hunter} R.~B.,  1967, \mn@doi [\mnras] {10.1093/mnras/136.3.245}, \href
  {https://ui.adsabs.harvard.edu/abs/1967MNRAS.136..245H} {136, 245}

\bibitem[\protect\citeauthoryear{{Jewitt}}{{Jewitt}}{2012}]{ActiveA}
{Jewitt} D.,  2012, \mn@doi [\aj] {10.1088/0004-6256/143/3/66}, \href
  {https://ui.adsabs.harvard.edu/abs/2012AJ....143...66J} {143, 66}

\bibitem[\protect\citeauthoryear{{Jewitt} \& {Haghighipour}}{{Jewitt} \&
  {Haghighipour}}{2007}]{JH07}
{Jewitt} D.,  {Haghighipour} N.,  2007, \mn@doi [\araa]
  {10.1146/annurev.astro.44.051905.092459}, \href
  {https://ui.adsabs.harvard.edu/abs/2007ARA&A..45..261J} {45, 261}

\bibitem[\protect\citeauthoryear{{Kalas}, {Graham}  \& {Clampin}}{{Kalas}
  et~al.}{2005}]{Kalas05}
{Kalas} P.,  {Graham} J.~R.,   {Clampin} M.,  2005, \mn@doi [\nat]
  {10.1038/nature03601}, \href
  {https://ui.adsabs.harvard.edu/abs/2005Natur.435.1067K} {435, 1067}

\bibitem[\protect\citeauthoryear{{Kalas} et~al.,}{{Kalas}
  et~al.}{2008}]{Kalas08}
{Kalas} P.,  et~al., 2008, \mn@doi [Science] {10.1126/science.1166609}, \href
  {https://ui.adsabs.harvard.edu/abs/2008Sci...322.1345K} {322, 1345}

\bibitem[\protect\citeauthoryear{{Kalas}, {Graham}, {Fitzgerald}  \&
  {Clampin}}{{Kalas} et~al.}{2013}]{Kalas13}
{Kalas} P.,  {Graham} J.~R.,  {Fitzgerald} M.~P.,   {Clampin} M.,  2013,
  \mn@doi [\apj] {10.1088/0004-637X/775/1/56}, \href
  {https://ui.adsabs.harvard.edu/abs/2013ApJ...775...56K} {775, 56}

\bibitem[\protect\citeauthoryear{{Kanavos}, {Markellos}, {Perdios}  \&
  {Douskos}}{{Kanavos} et~al.}{2002}]{KMP02}
{Kanavos} S.~S.,  {Markellos} V.~V.,  {Perdios} E.~A.,   {Douskos} C.~N.,
  2002, \mn@doi [Earth Moon and Planets] {10.1023/A:1026238123759}, \href
  {https://ui.adsabs.harvard.edu/abs/2002EM&P...91..223K} {91, 223}

\bibitem[\protect\citeauthoryear{{Kennedy} \& {Wyatt}}{{Kennedy} \&
  {Wyatt}}{2011}]{KW11}
{Kennedy} G.~M.,  {Wyatt} M.~C.,  2011, \mn@doi [\mnras]
  {10.1111/j.1365-2966.2010.18041.x}, \href
  {https://ui.adsabs.harvard.edu/abs/2011MNRAS.412.2137K} {412, 2137}

\bibitem[\protect\citeauthoryear{{Kennedy}, {Lovell}, {Kalas}  \&
  {Fitzgerald}}{{Kennedy} et~al.}{2023}]{KLKF}
{Kennedy} G.~M.,  {Lovell} J.~B.,  {Kalas} P.,   {Fitzgerald} M.~P.,  2023,
  \mn@doi [arXiv e-prints] {10.48550/arXiv.2305.10480}, \href
  {https://ui.adsabs.harvard.edu/abs/2023arXiv230510480K} {p. arXiv:2305.10480}

\bibitem[\protect\citeauthoryear{{Kushvah}}{{Kushvah}}{2008}]{Kush08}
{Kushvah} B.~S.,  2008, \mn@doi [\apss] {10.1007/s10509-008-9823-6}, \href
  {https://ui.adsabs.harvard.edu/abs/2008Ap&SS.315..231K} {315, 231}

\bibitem[\protect\citeauthoryear{{Markellos}, {Roy}, {Velgakis}  \&
  {Kanavos}}{{Markellos} et~al.}{2000}]{MRV00}
{Markellos} V.~V.,  {Roy} A.~E.,  {Velgakis} M.~J.,   {Kanavos} S.~S.,  2000,
  \mn@doi [\apss] {10.1023/A:1002487228086}, \href
  {https://ui.adsabs.harvard.edu/abs/2000Ap&SS.271..293M} {271, 293}

\bibitem[\protect\citeauthoryear{{Mikkola} \& {Innanen}}{{Mikkola} \&
  {Innanen}}{1997}]{MI97}
{Mikkola} S.,  {Innanen} K.,  1997, in {Dvorak} R.,  {Henrard} J.,  eds, The
  Dynamical Behaviour of our Planetary System. p.~345

\bibitem[\protect\citeauthoryear{{Murray} \& {Dermott}}{{Murray} \&
  {Dermott}}{2000}]{MD2000}
{Murray} C.~D.,  {Dermott} S.~F.,  2000, {Solar System Dynamics},
  \mn@doi{10.1017/CBO9781139174817.
}

\bibitem[\protect\citeauthoryear{{Nesvorn{\'y}}, {Alvarellos}, {Dones}  \&
  {Levison}}{{Nesvorn{\'y}} et~al.}{2003}]{NADL03}
{Nesvorn{\'y}} D.,  {Alvarellos} J. L.~A.,  {Dones} L.,   {Levison} H.~F.,
  2003, \mn@doi [\aj] {10.1086/375461}, \href
  {https://ui.adsabs.harvard.edu/abs/2003AJ....126..398N} {126, 398}

\bibitem[\protect\citeauthoryear{{Perdiou}, {Perdios}  \&
  {Kalantonis}}{{Perdiou} et~al.}{2012}]{PPK12}
{Perdiou} A.~E.,  {Perdios} E.~A.,   {Kalantonis} V.~S.,  2012, \mn@doi [\apss]
  {10.1007/s10509-012-1145-z}, \href
  {https://ui.adsabs.harvard.edu/abs/2012Ap&SS.342...19P} {342, 19}

\bibitem[\protect\citeauthoryear{{Quillen}}{{Quillen}}{2006}]{Q06}
{Quillen} A.~C.,  2006, \mn@doi [\mnras] {10.1111/j.1745-3933.2006.00216.x},
  \href {https://ui.adsabs.harvard.edu/abs/2006MNRAS.372L..14Q} {372, L14}

\bibitem[\protect\citeauthoryear{{Radzievskii}}{{Radzievskii}}{1950}]{Rad50}
{Radzievskii} V.~V.,  1950, \azh, \href
  {https://ui.adsabs.harvard.edu/abs/inothtere} {27, 250}

\bibitem[\protect\citeauthoryear{{Radzievskii}}{{Radzievskii}}{1953}]{Rad53}
{Radzievskii} V.~V.,  1953, \azh, \href
  {https://ui.adsabs.harvard.edu/abs/inothtere} {30, 265}

\bibitem[\protect\citeauthoryear{{Schuerman}}{{Schuerman}}{1980}]{Sch80}
{Schuerman} D.~W.,  1980, \mn@doi [\apj] {10.1086/157989}, \href
  {https://ui.adsabs.harvard.edu/abs/1980ApJ...238..337S} {238, 337}

\bibitem[\protect\citeauthoryear{{Simmons}, {McDonald}  \& {Brown}}{{Simmons}
  et~al.}{1985}]{SMB85}
{Simmons} J.~F.~L.,  {McDonald} A.~J.~C.,   {Brown} J.~C.,  1985, \mn@doi
  [Celestial Mechanics] {10.1007/BF01227667}, \href
  {https://ui.adsabs.harvard.edu/abs/1985CeMec..35..145S} {35, 145}

\bibitem[\protect\citeauthoryear{{Tremaine}}{{Tremaine}}{2023}]{Tremaine23}
{Tremaine} S.,  2023, {Dynamics of Planetary Systems}

\bibitem[\protect\citeauthoryear{{Wiegert}, {Innanen}  \& {Mikkola}}{{Wiegert}
  et~al.}{2000}]{WIM00}
{Wiegert} P.,  {Innanen} K.,   {Mikkola} S.,  2000, \mn@doi [\aj]
  {10.1086/301291}, \href
  {https://ui.adsabs.harvard.edu/abs/2000AJ....119.1978W} {119, 1978}

\bibitem[\protect\citeauthoryear{{Zotos}}{{Zotos}}{2015}]{Zot15}
{Zotos} E.~E.,  2015, \mn@doi [\apss] {10.1007/s10509-015-2513-2}, \href
  {https://ui.adsabs.harvard.edu/abs/2015Ap&SS.360....1Z} {360, 1}

\makeatother
\end{thebibliography}

\appendix

\section{A: Orbit Gallery}
\label{Gallery}

The primary family of stable prograde orbits in Henon's analysis is $g$, intersecting with a secondary family $g'$ that emerges below
a critical value of $C_H$. The equivalent families $G$ and $G'$ for non-zero $\beta$ do not map directly onto $g$
and $g'$ because they do not intersect.

\begin{figure}
\centering
\includegraphics[width=0.8\linewidth]{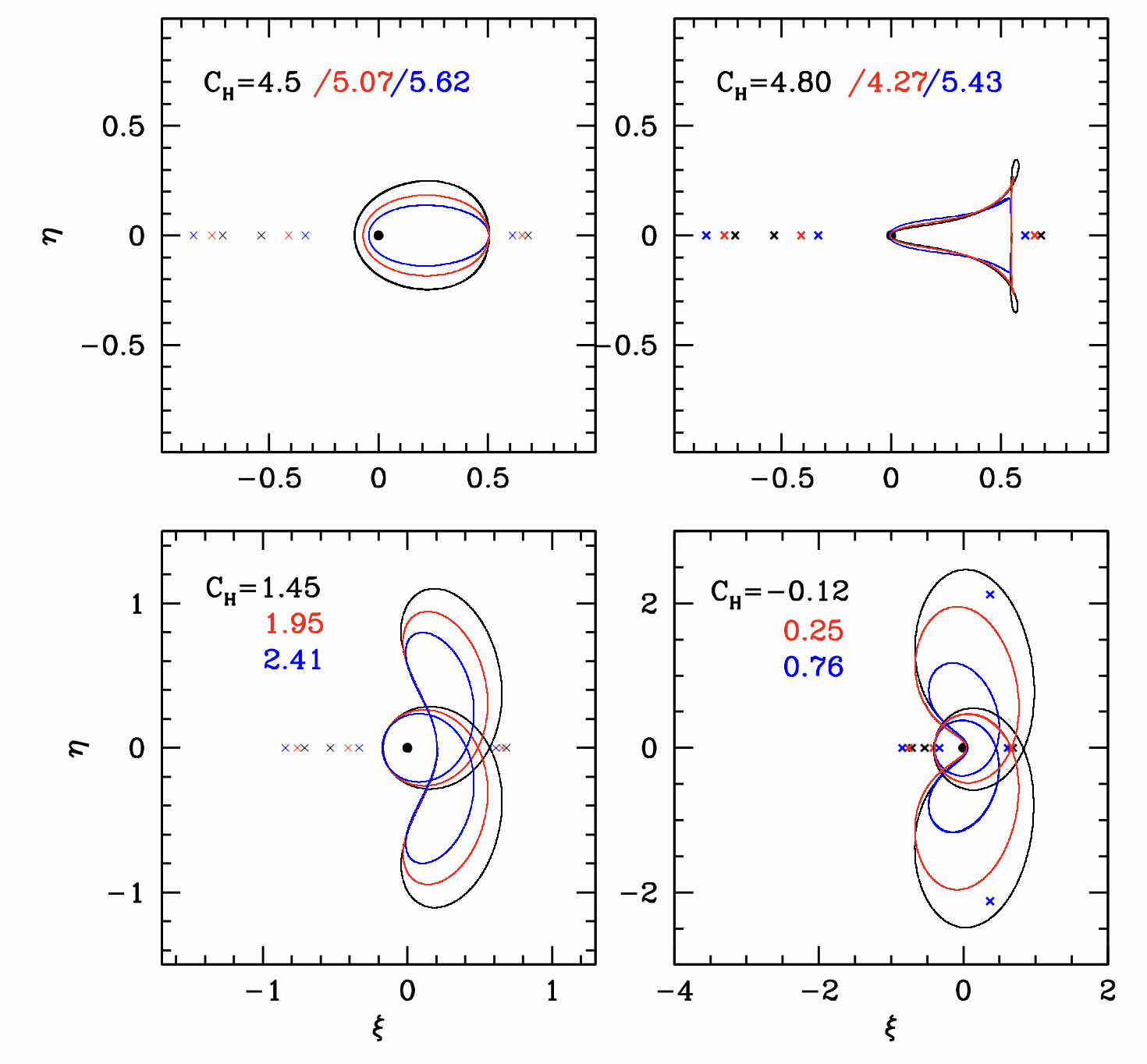}
\caption{These panels show how the shape of the $G$ family of orbits changes with $C_H$.
In the upper left panel, we show the equilibria with $\xi(0)=0.24$ for
$\beta=0.001$(black), $\beta=0.05$ (red) and $\beta=0.1$ (blue). Also shown, as crosses, are the
corresponding $L_1$,$L_2$ and $L_2'$ points for each $\beta$.  The upper right panel shows the
equivalent for  orbits close to the widest point of the stable branch in Figure~\ref{Orbit1}. 
The lower left panel shows the shapes of the orbits after the family has
become doubly periodic, and the lower right panel shows the continued evolution towards lower
$C_H$.
 \label{FamilyG}}
\end{figure}

 Figure~\ref{FamilyG} shows
the generalisation $G$ of this family ($g$) for $\mu=0.001$ and $\beta=0.001$ (black), $\beta=0.05$ (red) and
$\beta=0.1$ (blue). Each panel is chosen to have the same $\xi(0)$, which means that the $C_H$ will
be different for different $\beta$. The progression from the upper left to bottom right follows the decrease
of $C_H$ (for fixed $\beta$). The principal qualitative effect of increasing $\beta$ is compression of
the orbit shape in the $\eta$ direction, i.e. the effect of radiation pressure is to elongate the orbit
along the primary--secondary axis.

We have designated this family as $G$ because we wish to use this for the main family of stable prograde
orbits. However, the comparison with Figure~\ref{FamilyG} and the corresponding Figures~4, 5 and 6 of
\cite{Hen69} shows that the orbits for lower $C_H$ represent the generalisation of $g'$ rather than $g$. 
Note also that, in the lower two panels, the solutions are double periodic and include an intersection
moving in the positive direction at $\xi < 0$. These are the origin of the $G$ curves in Figure~\ref{Retro}.

\begin{figure}
\centering
\includegraphics[width=0.8\linewidth]{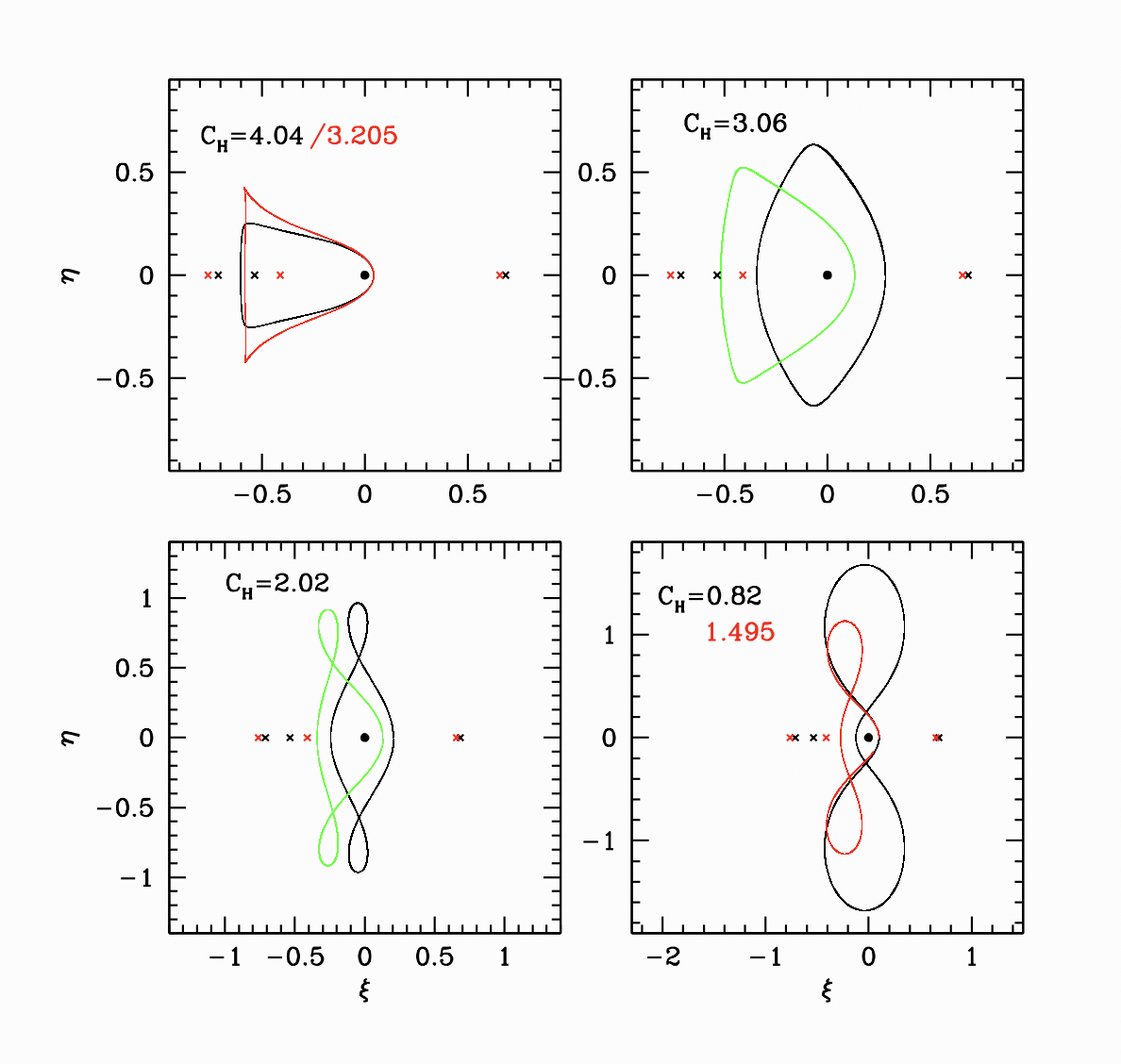}
\caption{ These panels show the evolution of the $G'$ family of orbits.
In the upper left panel, we show  orbits for $\beta=0.001$ (black) and $\beta=0.05$ (red). The
linear equilibrium points are shown as crosses.The two solutions have the same starting value $\xi(0)$. In the upper
right and lower left, there are no solutions for $\beta=0.05$ with equivalent $\xi(0)$. In these cases, orbits with the
same $C_H$ are shown as green curves. Finally, in the lower right, we show solutions for both $\beta$ and the
same $\xi(0)$.
 \label{FamilyGp}}
\end{figure}

Figure~\ref{FamilyGp} shows the orbits of the $G'$ family, as a function of $C_H$. This is composed of 
a combination of $g$ and $g'$ from the $\beta=0$ case, although most of them correspond to the
$g$ family. The black curves show $\beta=0.001$ and the
red curves show the case for $\beta=0.05$. This family shrinks in extent as $\beta$ increases, and so
there are no red curves in the upper right and lower left panels, because there are no solutions for these
values of $\xi(0)$. Furthermore, there are no blue curves because this family is absent for $\beta=0.1$. 
We show green curves in those cases where there is no solution at $\beta=0.05$. In these cases, $C_H$ 
is chosen to have the same value for $\beta=0.001$ and $\beta=0.05$.

\begin{figure}
\centering
\includegraphics[width=0.8\linewidth]{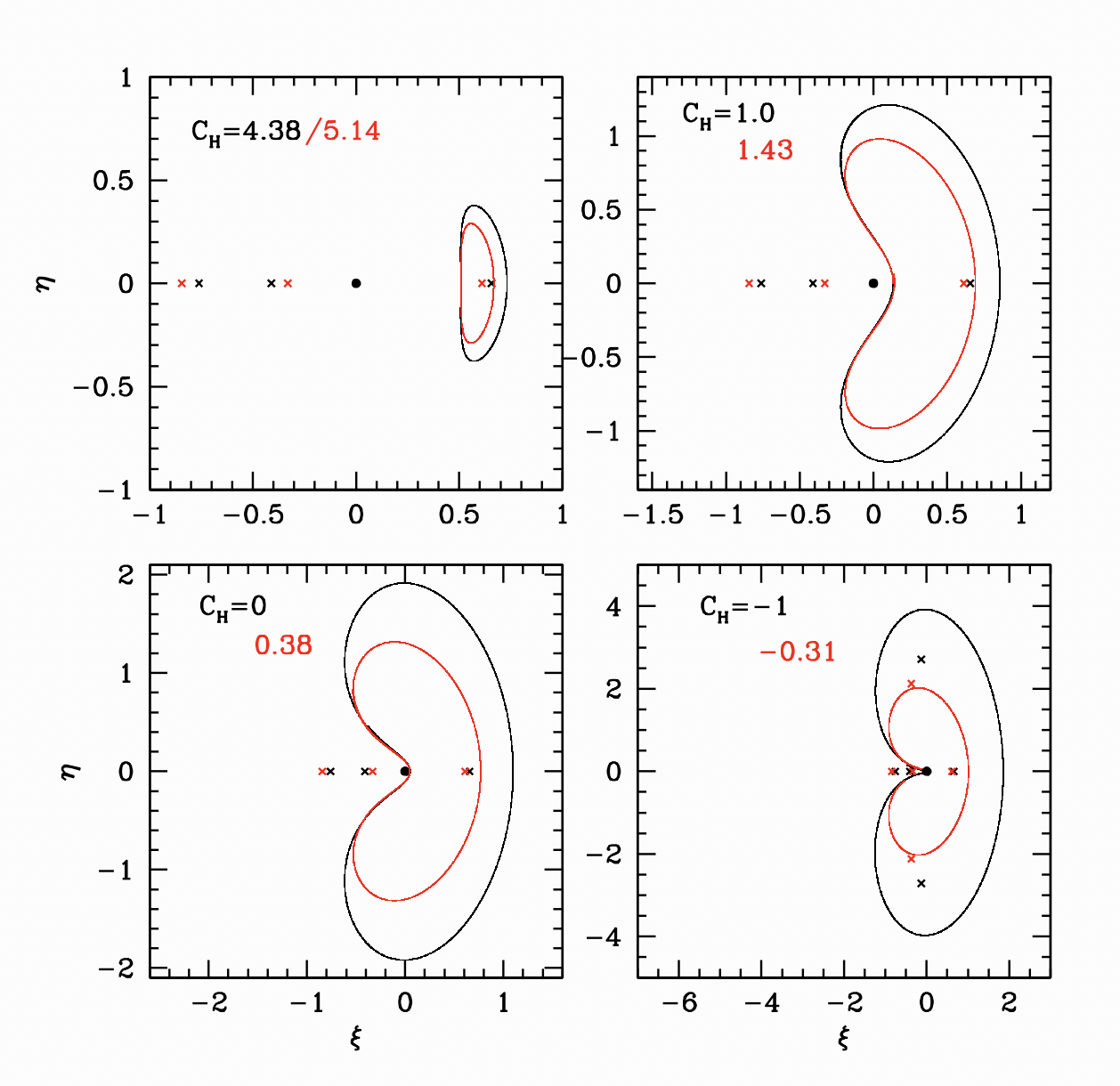}
\caption{In each panel, the black orbits refer to the case of $\beta=0.05$ and the red orbits for $\beta=0.1$. The orbits correspond to the family $A$ of equilibria. 
The crosses show the locations of the respective fixed points. In particular, this family corresponds to the
librations about the $L_2$ point.  As $C_H$ decreases (from upper left towards lower right) the orbits acquire larger amplitudes and become more distorted. \label{FamilyA}}
\end{figure}

To complete the discussion of prograde orbital families, Figure~\ref{FamilyA} shows the evolution of 
family $A$ (the orbits of libration about the $L_2$ point), for different $C_H$ and $\beta=0.05$ and $\beta=0.1$.
Once again, the comparison in each panel is made with fixed $\xi(0)$. We see that increasing $\beta$ makes the
orbit more compact.  The general shape trend is similar to that in the $\beta=0$ case, with the orbits tending
towards the consecutive collision shape as $C_H$ becomes more negative.

\begin{figure}
\centering
\includegraphics[width=0.8\linewidth]{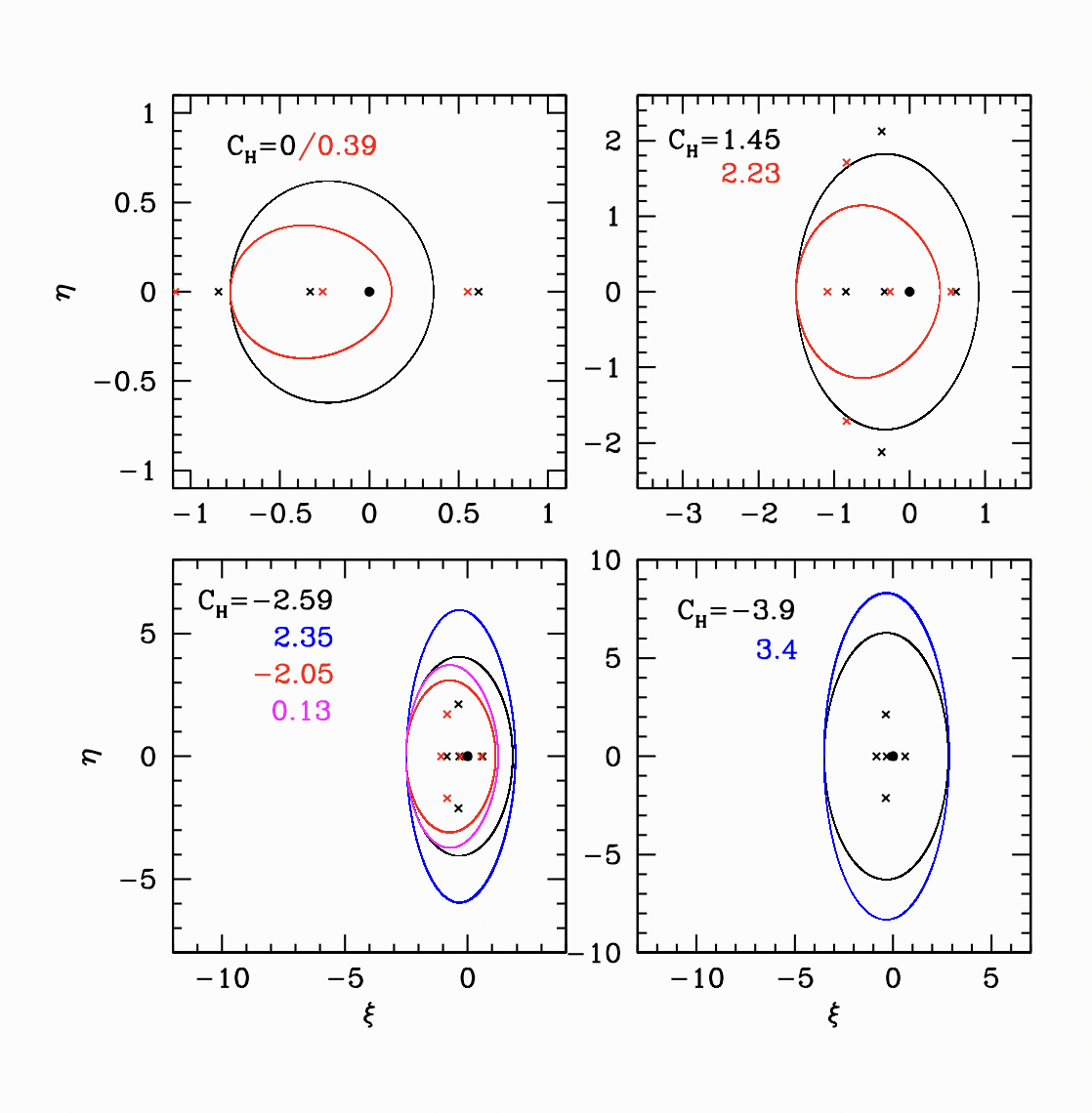}
\caption{These panels correspond to the famiily $F$ of retrograde orbits.
In each panel, the black orbits refer to the case of $\beta=0.1$ and the red orbits for $\beta=0.2$. The crosses show the locations of the respective fixed points. The blue curves represent the $F'$ family of the $\beta=0.1$ case.  The
magenta curve represents the $F'$ solution in the case $\beta=0.2$. Once again, the overall trend with increasing $\beta$ is to make the orbits more compact. \label{FamilyF}}
\end{figure}

The most important retrograde orbital family is $F$, the generalisation of Henon's $f$ family. Figure~\ref{FamilyF} shows
the evolution of this family as $C_H$ decreases, for $\beta=0.1$ and $\beta=0.2$. As seen previously in Figure~\ref{Retro},
the extension of $F$ to arbitrarily large distances remains for $\beta<0.11$, but fails for larger values. This is why there
is no red orbit in the bottom right panel of Figure~\ref{FamilyF}.

Figure~\ref{FamilyF} also demonstrates the appearance of the second branch of the $F$ family, which we have called
$F'$. This is represented by the second asymptotic solution discussed in \S~\ref{AsySol}, and is shown by the blue
curves. The orbital shape is similar but the extension is more pronounced in the $\eta$ direction.

\begin{figure}
\centering
\includegraphics[width=0.8\linewidth]{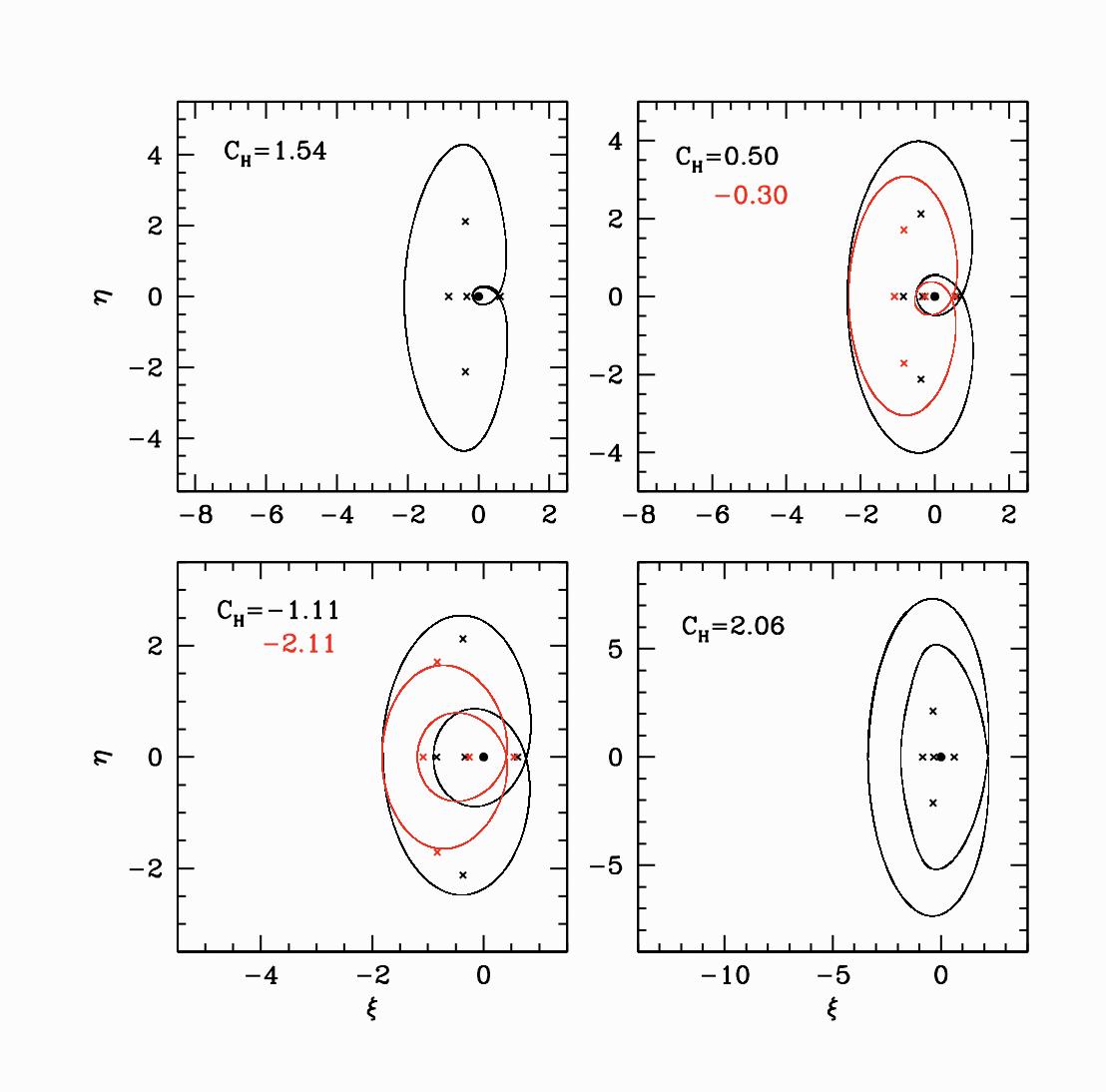}
\caption{In each panel, the black orbits refer to the case of $\beta=0.1$ and the red orbits for $\beta=0.2$. The crosses show the locations of the respective fixed points. The blue curves represent the $F'$ family of the $\beta=0.1$ case.  Once again, the overall trend with increasing $\beta$ is to make the orbits more compact. \label{FamilyGpp}}
\end{figure}

The $G$ curves in Figure~\ref{Retro} are simply the second axis crossing of those shown in Figure~\ref{FamilyGp}, but
the $\beta>0$ cases also show a doubly periodic family $G''$ with both crossings at $\xi<0$. The shape of this family
is shown in Figure~\ref{FamilyGpp}. This family is responsible for a lot of the structure in the middle and lower panels
of Figure~\ref{Retro}. The $\beta=0.1$ case shows how the two loops converge as the solution approaches the
turning point in $C_H$, and also how a second branch emerges to track the $F'$ family as the doubly periodic version.
The $\beta=0.2$ case does not extend over as large a range. Once again, we compare orbits with the same initial $\xi(0)$
and choose $C_H$ accordingly.

\begin{figure}
\centering
\includegraphics[width=0.8\linewidth]{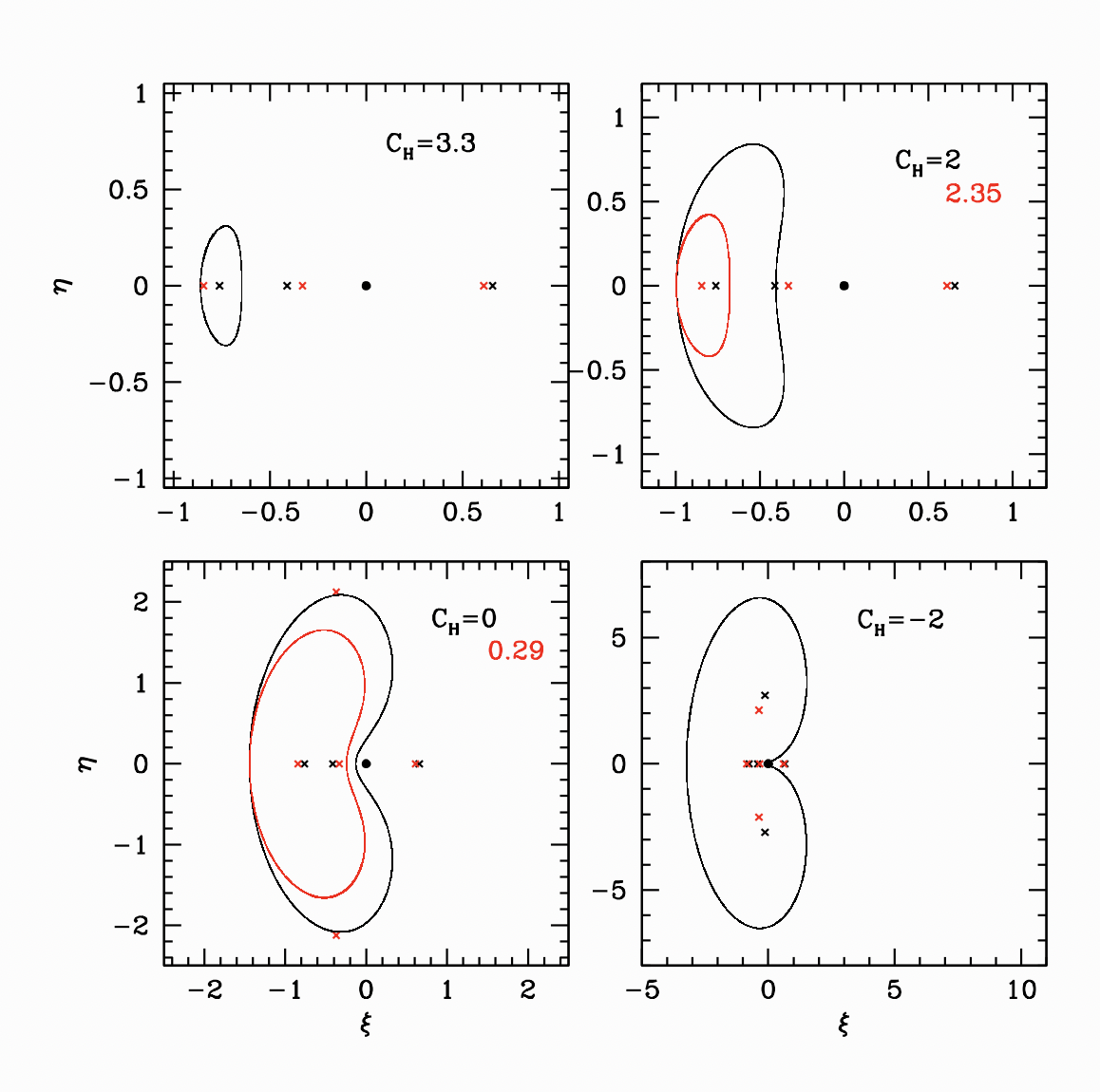}
\caption{The equilibria shown here represent the family $C$ of librations about the $L_1$ point. 
In each panel, the black orbits refer to the case of $\beta=0.05$ and the red orbits for $\beta=0.1$. The crosses show the locations of the respective fixed points. \label{FamilyC}}
\end{figure}

Figure~\ref{FamilyC} rounds out the census of orbit families, showing the Family $C$ of libration orbits about the
$L_1$ point. As can be seen in Figure~\ref{Retro}, for $\beta=0.1$ the family does not extend to arbitrarily large
distances, and so some of the panels show only the $\beta=0.05$ case.

\begin{figure}
\centering
\includegraphics[width=0.8\linewidth]{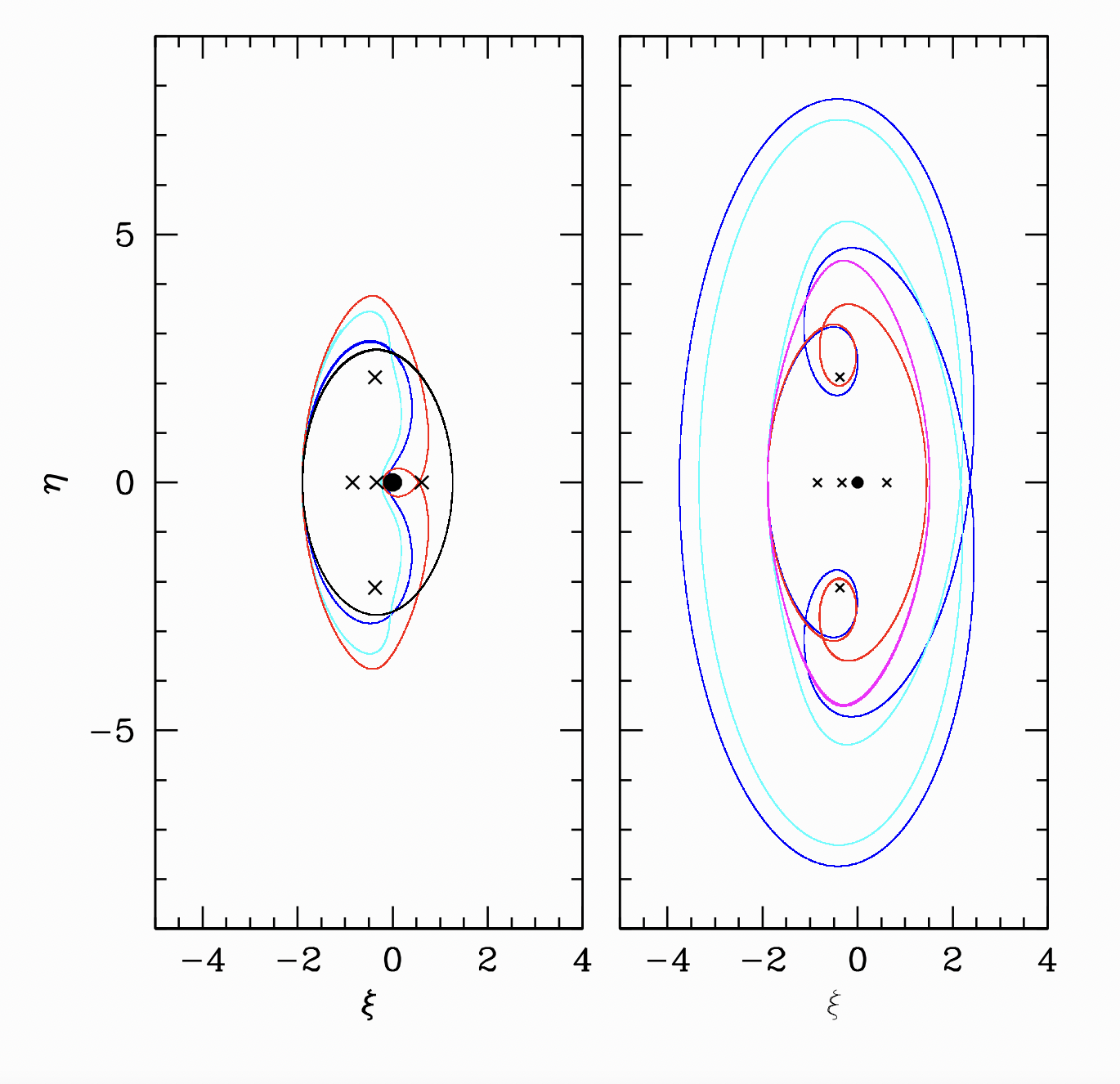}
\caption{All orbits in this figure are equilibria for the case $\mu=0.001$, $\beta=0.1$ and with starting condition
$\xi(0)=-1.9$ and $\dot{\xi}=0$. The value of $C_H$ is different, however. In the left hand panel, the black curve
represents the $F$ family, and the red curve indicates the leftmost intersection of the $G''$ family. Finally, the
blue and cyan curves represent two incarnations of the $C$ family. The right-hand panel shows two incarnations
each of the $G''$ family (blue and cyan) and the $F'$ family (red and magenta). Also show, as crosses, are the 
stationary points of the potential. The equivalents of the $L_4$ and $L_5$ points are responsible for the doubling
of the orbital families, as can be seen by the loops induced in the red and blue curves.
 \label{FamilyCon}}
\end{figure}

One of the more striking features of the $\beta=0.1$ and $\beta=0.2$ retrograde orbits is the `vortex-like' structure
apparent in the middle and lower panels of Figure~\ref{Retro}. To better understand the nature of this feature,
Figure~\ref{FamilyCon} shows the full set of retrograde equilibria that pass through $\xi(0)=-1.9$, for the case of
$\beta=0.1$. There are a total of eight different curves shown between the two panels. Some of the orbits correspond
to the standard incarnations to be expected (the $F$ orbit and the leftmost $G''$ orbit) but several of the orbital
families exhibit a doubling of the number of potential equilibria (this is responsible for the vortex-like structure
in Figure~\ref{Retro}).

We can trace the extra orbits to the appearance of the $L_4$/$L_5$ analogues (shown as crosses in Figure~\ref{FamilyCon})
because we note that several of the new orbits retain the overall shape of the parent family, but feature an extra loop
around the equilibrium point (compare the blue and cyan orbits in the right hand panel, or the red and magenta orbits).
So, we can trace the extra features in the orbital distribution to the presence of these potential minima, which affect the
orbital shape if the orbit passes too close to them.  It is important to note that these particular solutions within the
context of the Hill problem will
likely experience significant modification when translated to the full restricted three-body problem, as the positions of the
$L_4$ and $L_5$ equilibria will be affected by the inclusion of higher order terms truncated in the expansions used to define the Hill problem.
We have included them here for completeness, and note that they have little role in determining the potentially observable
features, as these orbits are unstable.

\section{B: Asymptotic Solutions}

\label{AsySol}

We can gain some preliminary insight into the influence of radiation pressure by considering the asymptotic
limits of the periodic orbits in the limit 
$\Delta \rightarrow \infty$, so we are looking
for simply periodic solutions to the equations
\begin{eqnarray}
\ddot{\xi} - 2\dot{\eta} & = & Q' + \left( 3 - 2\beta \right) \xi  \label{PhotAs1} \\
\ddot{\eta} + 2 \dot{\xi} & = & \beta \eta \label{PhotAs2}
\end{eqnarray}

We will adopt a trial solution of $\xi = \xi_0 + K_1 \cos \omega t$ and $\eta = K_2 \sin \omega t$ (assuming
a judiciously chosen origin for time). Note
that there is no term linear in $t$ as in Henon's version, because of the $\beta \eta$ term on the right
hand side of equation~(\ref{PhotAs2}) and 
 because the presence of the $Q'$ term is sufficient
to determine $\xi_0$ (which would require a contribution from the $\dot{\eta}$ term otherwise).  

\begin{figure}
    \centering
\includegraphics[width=1.0\linewidth]{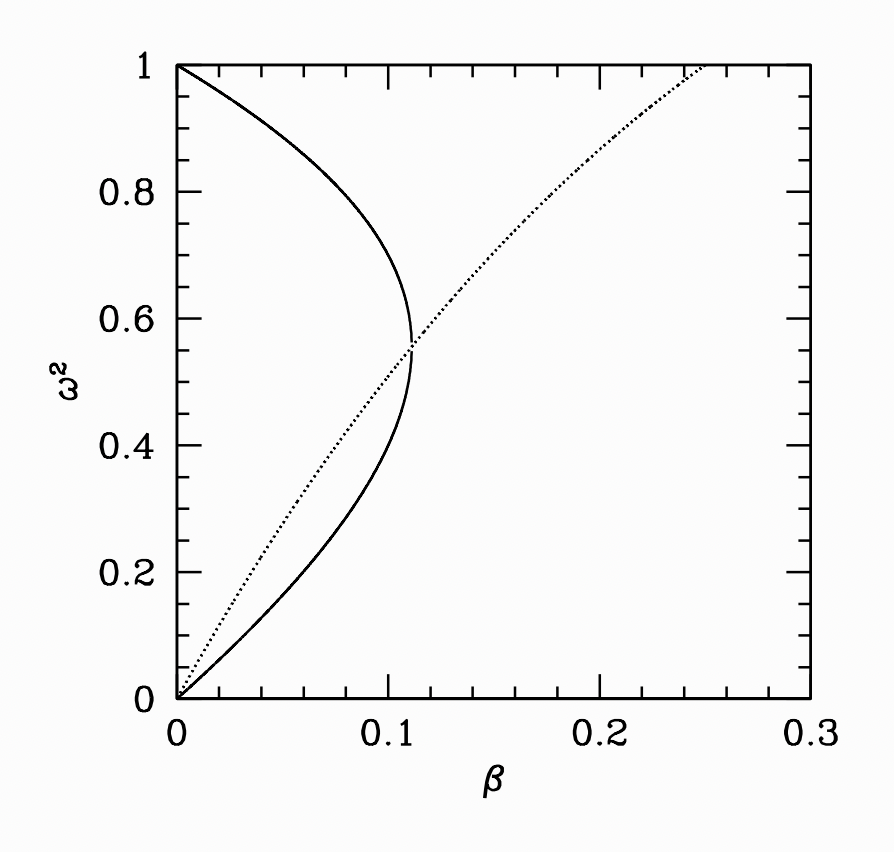}
\caption{The solid curve indicates the roots of equation~(\ref{dispeq}) for the asymptotic solution for $\mu=10^{-3}$. 
The dotted line represents the condition that the coefficient in front of $K_1$ in equation~(\ref{Cheq}) remains
positive (so that it still applies to the retrograde case).  Thus, the branch of the solid curve above the dotted line
corresponds to the standard asymptotic solution (which applies as $C_H \rightarrow - \infty$). The lower branch
represents a second asymptotic solution, which applies in the limit $C_H \rightarrow \infty$.
 \label{Disp}}
\end{figure}

Inserting this trial solution  into equations~(\ref{PhotAs1}) and (\ref{PhotAs2}) implies a requirement on the frequency, namely
\begin{equation}
4 \omega^2 = \left( \omega^2 + \beta \right) \left( \omega^2 + 3 - 2 \beta \right)
\end{equation}
which yields the solution
\begin{equation}
\omega^2 = \frac{1}{2} \left[ 1 + \beta \pm \left( 1 - 10 \beta + 9 \beta^2 \right)^{1/2} \right] \label{dispeq}
\end{equation}
For $\beta = 0$, the solutions are either $\omega^2=0$ or $\omega^2=1$ (Henon's solution). For finite
$\beta$, we have real  solutions for $\beta< \beta_{crit} = 0.11095$. Larger values yield complex frequencies, suggesting that
there are no stable asymptotic solutions for larger $\beta$. Figure~\ref{Disp} demonstrates the form of these
solutions.

\subsection{The quasi-satellites}

To complete the solution, $\xi_0 = -Q'/(3 - 2 \beta)$. The expression for the constant, in this limit, is
\begin{equation}
C_H = - \frac{(Q')^2}{3 - 2 \beta} - \frac{\omega^4 - \beta (3 - 2 \beta)}{\omega^2 + \beta} K_1^2 \label{Cheq}
\end{equation}
This recovers $C_H = - K_1^2$ in the $\beta \rightarrow 0$ limit, demonstrating that this is the asymptotic
form of Henon's family $f$. We note that the sign of the coefficient
to $K_1$ in equation~(\ref{Cheq}) changes when $\omega^2 = 2 \beta (3 - 2 \beta)/(1+\beta)$, which also happens to mark
the maximum $\beta$ at which $\omega$ is real. 

In the purely gravitational case, Henon showed that the retrograde family $f$ show stable equilibria in the
asymptotic limit of large distances, and we now understand that these are associated with the class of orbits termed `quasi-satellites' \citep{MI97,WIM00,GBM10}.  These are orbits whose heliocentric parameters share the same semi-major axis as the planet, but
which exhibit finite eccentricities, and which circulate about the planet, on scales larger than the Hill sphere, when viewed from the planetary rest frame -- as illustrated in Figure~\ref{3bEx}.
 Our results demonstrate that the existence of this class of orbits is not universal under the influence of radiation
pressure. For strong enough $\beta$ there is no longer such a stable asymptotic solution.

Furthermore, even for $\beta<\beta_{crit}$, we find two solutions for $\omega^2$. One applies in the
limit $C_H \rightarrow -\infty$ (as occurs in the purely gravitational case), and the other in the limit $C_H \rightarrow + \infty$.
This suggests that finite $\beta$ below the critical value should actually yield two equilibrium families at fixed $K_1$, rather than the single one in the $\beta=0$ limit. The solutions are given by
\begin{eqnarray}
\xi &=& - \frac{Q'}{3-2\beta}  + K_1 \cos \omega t \\
\eta & = & - \frac{2 \omega}{\omega^2 + \beta} K_1 \sin \omega t
\end{eqnarray}
Thus, both solutions correspond to ellipses, although the amplitude in the $\eta$ direction will be different at the same
$K_1$ (also corresponding to a different $C_H$). 

This second asymptotic solution does not occur in the Photogravitational Hill problem \citep{MRV00,KMP02,PPK12}, where
only the shift due to the $Q'$ is introduced. However, stable orbits corresponding to this equilibrium can be observed in
numerical orbital integrations of the full restricted three body problem with radiation \citep[e.g.][]{Zot15}. As we shall see below, this solution may also have observational implications for the properties of the enigmatic object Fomalhaut~b.

\subsection{Orbits with Consecutive Collisions}
\label{CC}

Henon also discussed a class of solutions wherein the orbits exhibited excursions to arbitrarily large distances, but also
returned periodically to $\xi=0$, $\eta=0$ to scatter off the secondary, terming these the ``orbits with consecutive collisions''.
These proved to be the asymptotic forms of the families $a$, $c$, $g$ and $g'$. However, an integral part of these
asymptotic solutions were terms that provided the $\eta$ component of the solution with either an additive constant or a
term linear in time. As noted above, equation~(\ref{PhotAs2}) no longer admits such solutions, and so we find no equivalents 
to these asymptotic solutions in this case. 

 

\section{C: Heliocentric Parameters}
\label{HelioFp}

In order to determine whether a directly imaged feature is bound to the host star, one measures the proper motion (and radial velocity, if one is lucky) relative to the star and converts the results into a set of Keplerian orbital parameters.  If the object is experiencing additional accelerations due to  an unseen companion, the resulting parameters may take values different from those expected given the position relative to the star.

\begin{figure}
\centering
\includegraphics[width=0.8\linewidth]{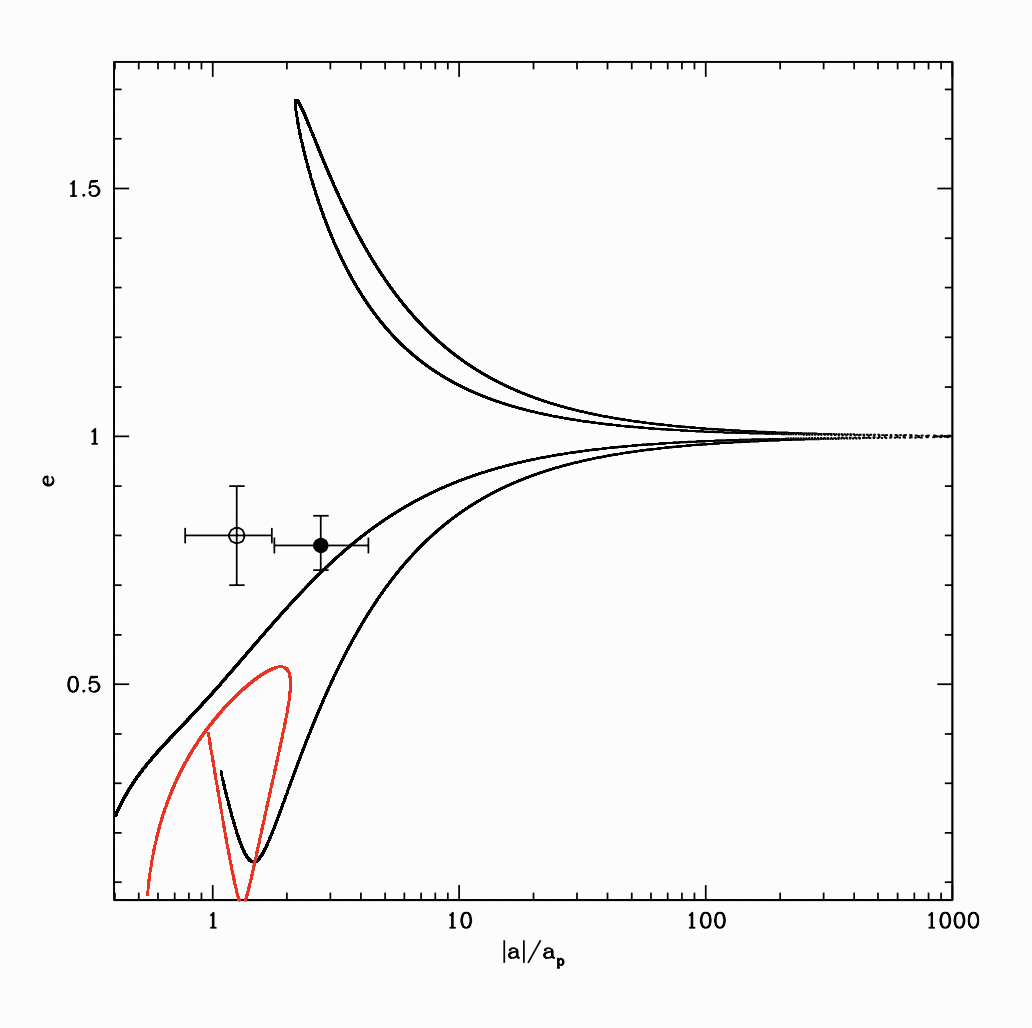}
\caption{The red curve shows the inferred heliocentric parameters for a quasi-satellite orbit drawn from the $F$ family,
in the case $\mu=0.001$ and $\beta=0.1$.  The initial conditions were $C_H=-5$ and $\xi(0)=-4.1455$. The black curve
shows the equivalent for a member of the $F'$ orbital family, with $C_H=5.85$ and $\xi(0)=-5$. The solid point represents the Fom~b parameters estimated by Gaspar \& Riecke (2020) 
and the open point those from Kalas et al. (2013). Semi-major axis is measured relative to the planet (which is taken to be the dust ring median for the observations).  We note again the black curve is a bound equilibrium simply periodic orbit -- the unbound heliocentric parameters are the result of accelerations due to the planet and not modelled in the Keplerian fit.
 \label{HelioF}}
\end{figure}

Figure~\ref{HelioF} shows the calculated heliocentric Keplerian semi-major axis and eccentricity for two example quasi-satellite orbits -- one from the $F$ family (shown in red) and one from the $F'$ (shown in black). These are both examples
of stable quasi-satellite orbits for the case of $\mu=0.001$ and $\beta=0.1$. We see that, during the wide excursions relative
to the planet, the inferred heliocentric parameters can vary wildly, including reaching apparently unbound values in the case
of the $F'$ orbit. We also compare this to the values inferred for the dusty object Fomalhaut~b \cite{Kalas13,GR20}. The
inferred orbital parameters of this object were the result of much discussion in the past, but are quite consistent with dust
in a quasi-satellite orbit. We discuss this more directly in \S~\ref{QuasiFom}.

\end{document}